\documentclass[acmsmall,screen]{acmart}
\AtBeginDocument{%
  \providecommand\BibTeX{{%
    Bib\TeX}}}
\definecolor{MyHighlight}{gray}{0.92}
\usepackage[]{collab}
\usepackage{xcolor}
\usepackage{colortbl}
\usepackage{tcolorbox}
\usepackage{mdframed}
\usepackage{listings}
\usepackage{makecell}
\usepackage{multirow}
\usepackage{array}

\usepackage{amsmath,amssymb,amsfonts}
\usepackage{algorithmic}
\usepackage{graphicx}
\usepackage{fbox}
\usepackage{textcomp}
\usepackage{minibox}
\usepackage{balance}
\usepackage{enumitem}
\usepackage{xspace}
\usepackage[T1]{fontenc}
\usepackage{booktabs}
\usepackage{threeparttable}
\usepackage{url}
\usepackage{soul}
\usepackage{ulem}
\usepackage{balance}
\usepackage{subfigure}
\usepackage{diagbox}
\usepackage{pifont}

\def\BibTeX{{\rm B\kern-.05em{\sc i\kern-.025em b}\kern-.08em
    T\kern-.1667em\lower.7ex\hbox{E}\kern-.125emX}}

\definecolor{ballblue}{rgb}{0.13, 0.67, 0.8}
\definecolor{grey}{rgb}{0.9, 0.9, 0.9}
\definecolor{googlered}{rgb}{0.914, 0.262, 0.207}
\definecolor{dandelion}{rgb}{0.95, 0.65, 0.0}
\definecolor{citecolor}{RGB}{106, 34, 107}

\definecolor{ballblue}{rgb}{0.13, 0.67, 0.8}
\definecolor{jcpink}{RGB}{255, 0, 96}
\collabAuthor{gb}{orange}{Guangba}
\collabAuthor{zr}{teal}{Zirui}
\collabAuthor{yl}{red}{Yulun}
\collabAuthor{ry}{dandelion}{Renyi}
\collabAuthor{yd}{orange}{Yuedong}

\definecolor{mygreen}{HTML}{AFCFA5}
\newcounter{summary}

\definecolor{mygreen}{HTML}{AFCFA5}
\newcounter{implication}

\definecolor{myyellow}{HTML}{FFF2CC}
\newcounter{finding}

\usepackage{adjustbox} 
\usepackage{booktabs}  
\usepackage{multirow}  


\newcommand\benchx{Cloud-OpsBench\xspace}
\newcommand\bench{Cloud-OpsBench}
\newcommand\num{754\xspace}
\newcommand\faulttype{57\xspace}

\begin{document}

\title{Cloud-OpsBench: A Reproducible Benchmark for Agentic Root Cause Analysis in Cloud Systems}

\author{Yilun Wang}
\orcid{0000-0001-9262-8652}
\email{ylwang25@cse.cuhk.edu.hk}
\affiliation{%
  \institution{The Chinese University of Hong Kong}
  \city{Hong Kong SAR}
  \country{China}
}
\author{Guangba Yu}
\orcid{0000-0001-6195-9088}
\email{guangbayu@cuhk.edu.hk}
\affiliation{%
  \institution{The Chinese University of Hong Kong}
  \city{Hong Kong SAR}
  \country{China}
}
\author{Haiyu Huang}
\orcid{0000-0001-6195-9088}
\email{hyhuang25@cse.cuhk.edu.hk}
\affiliation{%
  \institution{The Chinese University of Hong Kong}
  \city{Hong Kong SAR}
  \country{China}
}
\author{Yujie Huang}
\orcid{0000-0001-6195-9088}
\email{yjhuang@cse.cuhk.edu.hk}
\affiliation{%
  \institution{The Chinese University of Hong Kong}
  \city{Hong Kong SAR}
  \country{China}
}
\author{Zirui Wang}
\email{zrwang@cse.cuhk.edu.hk}
\orcid{0009-0004-5773-8716}
\affiliation{%
  \institution{The Chinese University of Hong Kong}
  \city{Hong Kong SAR}
  \country{China}
}

\author{Pengfei Chen}
\orcid{0000-0003-0972-6900}
\email{chenpf7@mail.sysu.edu.cn}
\affiliation{%
  \institution{Sun Yat-sen University}
  \city{Guangzhou City}
  \country{China}
}
\author{Michael R. Lyu}
\orcid{0000-0002-3666-5798}
\email{lyu@cse.cuhk.edu.hk}
\affiliation{%
  \institution{The Chinese University of Hong Kong}
  \city{Hong Kong SAR}
  \country{China}
}

\renewcommand{\shortauthors}{Yilun Wang, Guangba Yu et al.}

\begin{abstract}

LLM agents are explored for automating root cause analysis (RCA) in cloud-native systems, creating a need to evaluate both diagnostic correctness and the quality of the supporting investigation. Existing static benchmarks offer repeatable inputs but limited system-facing interaction, while live testbeds expose tools but hinder controlled comparison because incident evidence varies across runs; both paradigms focus primarily on final answers. To address these limitations, we present Cloud-OpsBench, an evaluation infrastructure for interactive and evidence-grounded cloud RCA. It comprises 754 runtime-verified cases across 57 fault types on two microservice workloads spanning application services and Kubernetes platform layers. Each fault is captured as a state snapshot and replayed through standard diagnostic interfaces, with outcome labels and diagnostic evidence graphs that enable matched comparisons and process-level analysis. Evaluating eleven LLMs on Cloud-OpsBench reveals that strong outcome accuracy often masks incomplete investigations: while Joint RCA Accuracy (JRA) reaches up to 0.78 on OnlineBoutique and 0.68 on TrainTicket, models identify the correct root cause without achieving evidence closure (e.g., in 33.1\% and 29.4\% of evaluated cases for GPT-5). These results demonstrate that final-answer accuracy alone cannot establish whether an agent has traced the failure and collected the evidence needed to confirm its root cause.
\end{abstract}






\maketitle

\section{Introduction}


Cloud-native systems host critical online services, but diagnosing their failures remains difficult because observable symptoms can propagate across application and platform layers~\cite{yu2020microscaler2,Microservie2018TSE,li2022incident,incident2020}.
Root-cause analysis is therefore a deployed-software debugging task that spans application code, deployment specifications, runtime configurations, service dependencies, and infrastructure states.
LLM-based agents offer a new approach to automating this task by selecting diagnostic tools and targets, interpreting returned observations, refining hypotheses, and consulting operational knowledge~\cite{RCACopilot,RCAgent2024CIKM,ewaschuk2016monitoring,jones2016troubleshooting}.
This shift changes the evaluation target from root-cause prediction over supplied telemetry to an interactive investigation whose conclusion must be supported by acquired observations.


Existing RCA benchmarks do not jointly support the conditions needed to evaluate this interactive investigation.
Static telemetry benchmarks provide fixed incident artifacts and reproducible inputs, but they do not test whether an agent can select diagnostic tools and targets based on observations returned by the system~\cite{Loghub2023ISSRE,RCA2025Eval,TimeSeriesBench2024ISSRE}.
Live-environment benchmarks preserve system-facing interaction, but the same injected fault may differ in severity, duration, or affected scope across runs~\cite{chen2025aiopslab,ITBench2025ICML,AIOpsArena25SANER}.
As a result, two agents evaluated in separate live runs may face different fault conditions, so their performance differences cannot be attributed solely to the agents.
Across both paradigms, outcome-oriented metrics evaluate the final diagnosis but cannot determine whether the agent has collected observations that causally justify its conclusion.
Consequently, existing evaluations only partially measure interactive evidence acquisition, run-to-run comparability, and diagnostic process quality.


To address these gaps, we present \benchx \footnote{https://github.com/LLM4Ops/Cloud-OpsBench}, an evaluation infrastructure for interactive and evidence-grounded cloud RCA.
It comprises \num fault cases across \faulttype fault types that span application services and Kubernetes platform layers, with OnlineBoutique and TrainTicket serving as representative workloads.
Each fault is realized and runtime-verified in a live Kubernetes testbed before its verified incident state and associated temporal telemetry are captured as a state snapshot.
During evaluation, snapshot-backed diagnostic interfaces replay the captured responses, allowing different agents to investigate the same realized incident while independently choosing their tools and targets.
Each case pairs a root-cause label with a milestone-based evidence graph, enabling evaluation of root-cause correctness, milestone coverage, evidence-order consistency, and investigation efficiency.



We evaluate eleven LLMs on \benchx under a common ReAct scaffold and further examine experience reuse and alternative agent architectures and harnesses.
First, diagnostic accuracy varies across models and fault types.
Across the eleven models, mean Joint RCA Accuracy (JRA) is 0.59 on OnlineBoutique and 0.49 on TrainTicket, with the highest scores reaching 0.78 and 0.68, respectively.
Performance- and code-related faults remain particularly challenging.
Second, correct diagnoses can lack complete supporting evidence.
For example, GPT-5 produces correct diagnoses with incomplete evidence in 33.1\% of evaluated cases on OnlineBoutique and 29.4\% on TrainTicket. Trajectory inspection shows that agents can submit the correct RCA result without tracing the failure to the affected backend or performing the checks needed to confirm the root cause.
Third, operational experience can improve diagnosis, but its benefits vary across models, systems, and reuse methods.
Fourth, with GLM-5.1 held fixed, different agent architectures and harnesses achieve similar diagnostic accuracy with substantially different evidence closure rates and token costs.
Higher computational expenditure does not consistently yield more accurate or more complete diagnoses.

\begin{table*}[t]
\centering
\caption{Comparison of \bench{} with existing AIOps benchmarks across the three evaluation gaps.}
\label{tab:comparison}
\begin{threeparttable}
    \footnotesize
        \begin{tabular*}{\textwidth}{@{\extracolsep{\fill}}llccc@{}}
        \toprule
        \textbf{Benchmark} & \textbf{Paradigm} & \makecell{\textbf{System Interaction}\\\textit{(Gap 1)}} & \makecell{\textbf{Run-to-Run Consistency}\\\textit{(Gap 2)}} & \makecell{\textbf{Process Evaluation}\\\textit{(Gap 3)}} \\
        \midrule
        \multicolumn{5}{@{}l}{\textit{\textbf{Static Telemetry Benchmarks}}} \\
        LogHub~\cite{Loghub2023ISSRE} & Static Artifact & \ding{55} & \ding{51} & \ding{55} \\
        RCAEval~\cite{RCA2025Eval} & Static Artifact & \ding{55} & \ding{51} & \ding{55} \\
        TimeSeriesBench~\cite{TimeSeriesBench2024ISSRE} & Static Artifact & \ding{55} & \ding{51} & \ding{55} \\
        Nezha~\cite{nezha2023fse} & Static Artifact & \ding{55} & \ding{51} & \ding{55} \\
        OpenRCA~\cite{OpenRCA2025ICLR} & Static Artifact & \ding{55}\tnote{1} & \ding{51} & \ding{55} \\
        \midrule
        \multicolumn{5}{@{}l}{\textit{\textbf{Dynamic Environment Benchmarks}}} \\
        AIOpsArena~\cite{AIOpsArena25SANER} & Live Environment & \ding{55} & \ding{55} & \ding{55} \\
        AIOpsLab~\cite{chen2025aiopslab} & Live Environment & \ding{51} & \ding{55} & \ding{55} \\
        ITBench~\cite{ITBench2025ICML} & Live Environment & \ding{51} & \ding{55} & \ding{55} \\
        SREGym~\cite{clark2026sregym} & Live Environment& \ding{51} & \ding{55} & \ding{55} \\
        ORCA-bench~\cite{gong2026orca} & Live Environment& \ding{51} & \ding{51}\tnote{2} & \ding{55} \\
        \midrule
        \multicolumn{5}{@{}l}{\textit{\textbf{Proposed Benchmark}}} \\
        \textbf{\benchx} & \textbf{State Snapshot} & \ding{51}\tnote{3} & \ding{51} & \ding{51} \\
        \bottomrule
        \end{tabular*}
        \begin{tablenotes}
            \scriptsize
            \item[1] Code execution over supplied offline artifacts is not counted as system-facing interaction because it does not query an operational interface.
            \item[2] It queries a fixed six-day historical telemetry data.
            \item[3] \bench{} replays captured cluster observations through snapshot-backed diagnostic interfaces for supported tool queries.
        \end{tablenotes}
\end{threeparttable}
\end{table*}

In summary, this paper makes the following contributions:
\begin{itemize}[leftmargin=*]

\item \textbf{A Runtime-Verified Fault Benchmark.}
We construct \benchx with \num fault cases spanning \faulttype fault types
across application services and Kubernetes platform layers on two
microservice workloads.
A knowledge-guided pipeline converts fault specifications into executable
injections and retains each case only after live runtime verification.

\item \textbf{Decoupled Fault Realization and Interactive Evaluation.}
We separate live fault realization from agent evaluation by capturing the
tool-observable evidence of each verified incident as a state snapshot.
Snapshot-backed interfaces then replay the same incident under matched
observations while allowing agents to choose diagnostic tools and targets
independently.

\item \textbf{Process-Grounded Diagnostic Evaluation.}
We represent diagnostic processes using milestone-based evidence graphs that encode completion logic, prerequisite relations, and alternative admissible evidence bundles. We define metrics for milestone coverage, evidence closure, evidence-order consistency, and investigation efficiency, without requiring a single prescribed expert command sequence.

\item \textbf{Experiments on LLM-based RCA Agents.}
We evaluate eleven LLMs on \benchx and reveal a gap between root-cause identification and diagnostic evidence closure: agents can correctly identify the faulty component and fault type without completing the supporting diagnostic evidence chain. These findings highlight that outcome accuracy alone is insufficient to establish the reliability of evidence-grounded diagnosis.

\end{itemize}

\section{Motivation}\label{sec:motivation}
\subsection{Agentic RCA}\label{sec:agenticrca}

Cloud-native root-cause analysis is a deployed-software debugging task.
In Kubernetes, software behavior is determined jointly by application code, deployment specifications, runtime configurations, service dependencies, and infrastructure states.
A failure reported at an entry service may therefore originate several components or system layers away.
Diagnosis must connect the reported symptom to a concrete faulty component and causal mechanism across these software and operational artifacts.
This paper focuses on this task in Kubernetes-based cloud-native systems.

Agentic RCA makes this debugging process interactive.
Given a user query describing an observed failure, an agent decides which system objects, logs, configurations, and dependency relations to inspect through diagnostic tools, and revises its hypotheses as observations accumulate~\cite{AidAI25FSE,StepFly2025}.
Consequently, two agents may produce the same root-cause label while differing substantially in whether their conclusions are supported by sufficient and causally coherent evidence.
A benchmark for agentic RCA must therefore evaluate both the diagnostic outcome and the evidence-acquisition process that supports it.

Agents may also consult operational knowledge, such as API semantics and
troubleshooting guidance, when interpreting observations~\cite{ewaschuk2016monitoring,jones2016troubleshooting,LearnKUBE}.
Such knowledge can be provided as a controlled experimental input, but any
accepted diagnosis must still be grounded in case-specific observations
returned by tools.


\subsection{Evaluation Gaps}



\begin{figure*}[t]
    \centering
    \includegraphics[width=\textwidth]{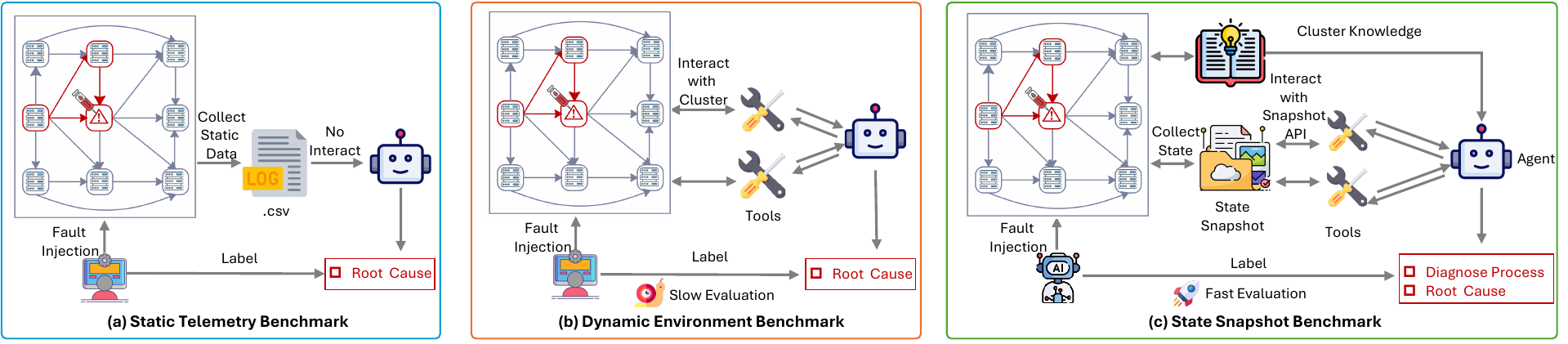}
    \vspace{-0.2in}
    \caption{Benchmarking paradigms for agentic RCA.
    (a)~Static telemetry: no system-facing interaction; outcome-only root-cause
    labels.
    (b)~Live environments: interactive diagnosis, but still outcome-only
    evaluation, with run-to-run variability and high cost.
    (c)~State snapshots (\benchx): matched interactive investigation over frozen
    incident evidence, with both process and outcome scoring.
    }
    \label{fig:motivation}
    \vspace{-0.2in}
\end{figure*}

Existing benchmarks leave complementary aspects of this investigation unmeasured.
Static benchmarks offer reproducible incident artifacts but do not evaluate interactive evidence collection.
Live-environment benchmarks support such interaction but expose agents to varying runtime observations, complicating process-level comparison.
Across both paradigms, outcome-oriented metrics cannot determine whether the collected evidence justifies the diagnosis.
Figure~\ref{fig:motivation} and Table~\ref{tab:comparison} summarize these limitations.

\subsubsection{Gap 1: Lack of Interactive Diagnosis in Static Benchmarks}
\label{sec:gap1}

Static RCA benchmarks typically provide pre-collected logs, metrics, or traces as fixed inputs~\cite{TimeSeriesBench2024ISSRE,Loghub2023ISSRE,RCA2025Eval,LogEval2025}.
This design supports reproducible analysis, and preprocessing or code-based filtering can reduce the amount of telemetry presented to a model.
However, it does not provide the system-facing interaction used in operational diagnosis, where an agent selects a diagnostic tool, specifies its target, and receives an observation before deciding what to inspect next.
Consequently, the benchmark evaluates reasoning over supplied evidence rather than the agent's ability to acquire evidence through an adaptive investigation.
Even settings that allow code execution over offline artifacts still operate on
a supplied file corpus rather than on system-facing diagnostic
interfaces~\cite{OpenRCA2025ICLR}.

In a static formulation of the TrainTicket DBReadOnly case, the model receives a fixed corpus of logs and telemetry and searches it offline for evidence such as read-only write rejections.
This setting evaluates evidence retrieval and interpretation within a predetermined corpus.
It does not evaluate whether, starting only from a query reporting a payment failure, the agent knows to query the dependencies of \texttt{ts-payment-service}, select \texttt{ts-order-service} from the returned topology, and request its logs.
The missing capability is therefore interactive evidence collection, in which the agent chooses what to query next based on observations returned by previous diagnostic actions.

\subsubsection{Gap 2: Variability Across Live Runs}
\label{sec:gap2}

Live-environment benchmarks deploy microservices and inject faults during evaluation, preserving system-facing interaction and realistic runtime behavior~\cite{chen2025aiopslab,ITBench2025ICML,AIOpsArena25SANER,clark2026sregym}.
Even when the fault type, target, and injection parameters are held constant, the resulting fault manifestation may vary in severity, duration, and affected scope across runs.
The realized impact is jointly determined by the fault injection and the system's transient runtime state.
Concurrent workloads, resource availability, scheduling decisions, and in-flight retries can amplify or attenuate the observable failure.

Such variation is valuable for evaluating robustness under realistic operational uncertainty.
However, it complicates controlled agent comparison.
When two agents are evaluated in separate live runs, both the diagnostic policy and the realized fault manifestation may differ.
Observed performance differences therefore cannot be attributed solely to the agents.
Repeated trials can estimate average performance and run-to-run variance, but do not guarantee identical tool observations across agents.

In addition, live evaluation incurs substantial lifecycle overhead.
Each independent run must establish a clean system state, warm up the workload, activate and verify the fault, allow the agent to complete its investigation, and restore the environment before the next trial.
Because reliable estimates require repeated runs for every combination of fault case, agent, and experimental configuration, this lifecycle cost grows multiplicatively with benchmark scale.
The resulting evaluation latency and infrastructure demand make large-scale comparison and rapid agent iteration difficult~\cite{MicroRemed}.

\subsubsection{Gap 3: Outcome-Only Evaluation}
\label{sec:gap3}

Existing RCA benchmarks primarily evaluate whether the predicted component and fault type match the ground truth using metrics such as top-$k$ accuracy, precision, recall, or F1-score~\cite{AIOpsArena25SANER,Loghub2023ISSRE,RCA2025Eval}.
These metrics remain necessary for measuring diagnostic correctness, but they provide little information about how an agent obtains its answer.
Two agents receive identical outcome credit when they return the same correct diagnosis, even if one has collected observations that connect the reported symptom to the root cause while the other submits the same diagnosis without collecting key supporting observations.
Conversely, an agent that conducts a largely grounded investigation but makes an incorrect final attribution is indistinguishable from one that fails to investigate the system.

This creates a gap in the construct validity of agentic RCA evaluation.
The claimed capability includes choosing diagnostic actions, acquiring relevant observations, and establishing a causal explanation, whereas outcome-only metrics measure only the final prediction.
Evaluation should therefore complement diagnostic accuracy with process measures over the observable tool-mediated investigation rather than the agent's self-reported reasoning~\cite{AI-NativeBench,AgentCompass}.
Such measures should assess the completeness, causal coherence, and efficiency of the investigation.

\subsection{Benchmark Requirements}
\label{sec:requirements}

The preceding gaps yield three requirements for evaluating agentic RCA.
\textbf{R1: System-facing evidence acquisition.}
An agent must be able to select tools and targets through a defined set of standard diagnostic interfaces, receive captured system observations, and adapt subsequent actions to the returned evidence.
\textbf{R2: Matched and reusable incident evidence.}
Different agents must investigate the same realized incident under the same available evidence and deterministic query-response mapping without redeploying or reinjecting the fault for each evaluation.
\textbf{R3: Process evaluation beyond a single reference trajectory.}
In addition to outcome correctness, the benchmark must assess the evidential completeness, causal order, and efficiency of the observable investigation without prescribing a canonical command sequence.
Evidence-equivalent diagnostic facts and independent investigation orders should receive equal credit.
We scope evaluation to interactive investigation over preserved,
runtime-verified incident evidence rather than to live remediation or
open-ended environment mutation.
These requirements guide the design of \benchx, introduced next.

\section{\benchx}\label{sec:method}

\begin{figure*}[t]
    \centering
    \includegraphics[width=\textwidth]{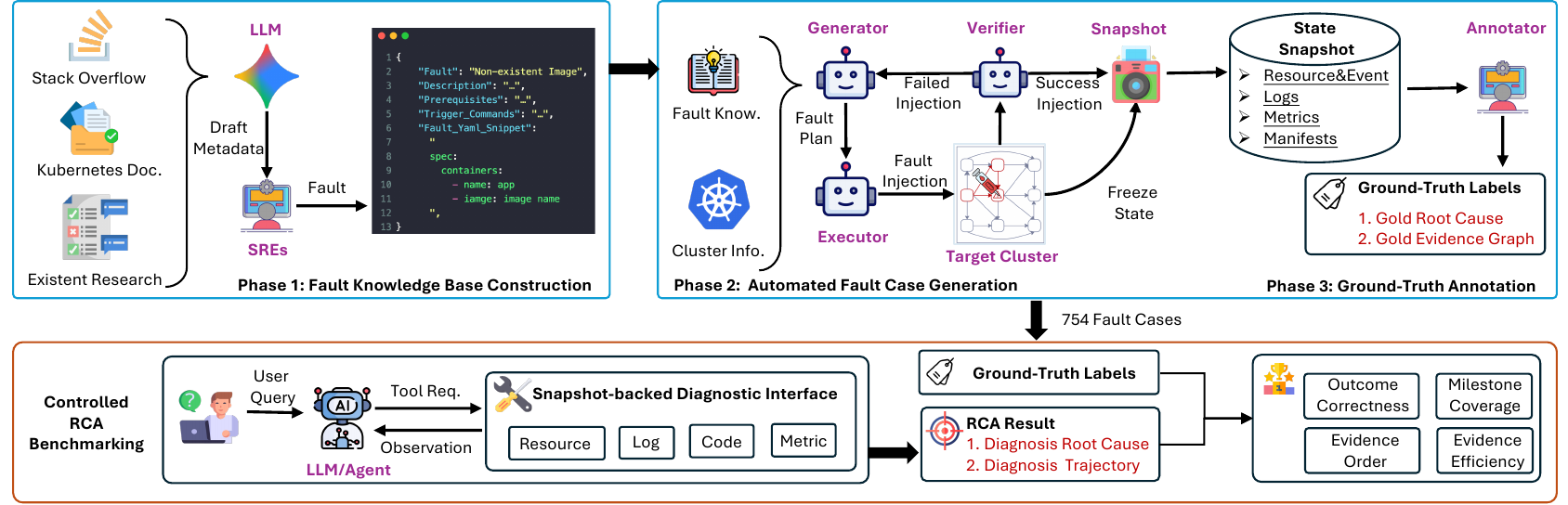}
    \vspace{-0.2in}
    \caption{Overview of \benchx: Phase~1 builds the fault knowledge base; Phase~2 instantiates, verifies, and snapshots each case; Phase~3 annotates outcome and process labels; agents then investigate through snapshot-backed diagnostic interfaces.}
    \label{fig:overview}
    \vspace{-0.1in}
\end{figure*}

To satisfy these requirements, we present \benchx, an evaluation infrastructure for interactive and evidence-grounded cloud RCA comprising \num fault cases across \faulttype fault types that span application services and Kubernetes platform layers.
As illustrated in Fig.~\ref{fig:overview}, \benchx realizes each fault in a live Kubernetes testbed through a knowledge-grounded construction pipeline (\S~\ref{sec:phase1}--\ref{sec:phase2}).
Snapshot-backed diagnostic interfaces satisfy R1 by supporting adaptive investigation through a defined diagnostic toolset (\S~\ref{sec:design_and_formulation}).
They satisfy R2 by replaying the same available incident evidence and query-response mapping without per-agent fault realization.
Each case carries a milestone-based evidence graph that satisfies R3 by evaluating established diagnostic facts and their prerequisite relations without requiring agents to follow a single reference command sequence (\S~\ref{sec:gt_graph}).

\subsection{Design Philosophy and Task Formulation}
\label{sec:design_and_formulation}



To bridge the tension between ecological validity and experimental control, we introduce the \textbf{State Snapshot} paradigm shown in Fig.~\ref{fig:motivation}(c).
Instead of evaluating agents directly against a live cluster, \benchx first realizes each fault in a real Kubernetes environment and captures its tool-observable context after the Verifier confirms the task-defining manifestation.
Each snapshot combines resource, configuration, topology, and component states collected after verification with timestamped events, logs, and metrics from the case-specific incident interval between pre-injection baseline confirmation and manifestation verification.
It therefore preserves both the verified incident state and its associated temporal telemetry without attempting to reproduce the complete system history.

During evaluation, agents interact with a snapshot-backed diagnostic interface that replays captured responses to supported tool queries, such as calling \texttt{GetResources} to inspect Pods.
This design does not eliminate noise from the original incident evidence.
Rather, it eliminates evaluation-time environmental nondeterminism: the snapshot preserves evidence produced by a real fault realization while fixing the observation space for agent comparisons.

Within this controlled replay environment, we formalize Agentic RCA as a trajectory-based decision process.
For each evaluation case $j$, the outcome ground truth is $\mathcal{R}_j^*=\langle C_j^*, F_j^* \rangle$, where $C_j^*$ denotes the faulty component and $F_j^*$ denotes the fault type.
Its process-level ground truth is represented by a diagnostic evidence graph $\mathcal{D}_j^*$, which specifies the intermediate diagnostic states required to justify $\mathcal{R}_j^*$ and is constructed in \S~\ref{sec:gt_graph}.
An evaluation episode is modeled as
   \vspace{-0.1in}
  \begin{equation}
      \Phi: \langle U_j, \mathcal{S}_j \rangle
      \rightarrow \langle \mathcal{T}_j, \hat{\mathcal{R}}_j \rangle .
  \end{equation}

The agent receives a user query $U_j$ describing the observed failure, diagnostic tool specifications, and the allowed component and fault-type labels.
Case-specific system observations are obtained through the diagnostic interface backed by the frozen snapshot $\mathcal{S}_j$, while the underlying snapshot repository is not directly exposed to the agent.
At each step, the agent issues a concrete tool use $t_i$, including the tool name and arguments, and the interface returns an observation $o_i$ replayed from $\mathcal{S}_j$.
The evaluation harness records the resulting trajectory $\mathcal{T}_j=[(t_1,o_1),\dots,(t_n,o_n)]$.
At termination, the agent returns only the predicted diagnosis $\hat{\mathcal{R}}_j=\langle \hat{C}_j, \hat{F}_j \rangle$.
We evaluate $\hat{\mathcal{R}}_j$ against $\mathcal{R}_j^*$ for outcome accuracy and $\mathcal{T}_j$ against $\mathcal{D}_j^*$ for process quality.





\subsection{Phase 1: Fault Knowledge Base Construction}\label{sec:phase1}
\benchx builds a fault taxonomy that spans both upper-layer microservices and the underlying infrastructure, synthesized from three complementary sources: (1) official Kubernetes documentation~\cite{k8sdoc} for specification-based misconfigurations (e.g., incorrect service selectors); (2) developer communities~\cite{stackoverflow,KodeKloud} for empirical operational faults; and (3) academic literature~\cite{Kubernetes2024DSN,chen2025aiopslab,ITBench2025ICML} for canonical failure modes.

Since raw sources combine unstructured prose with code snippets, we use Gemini 3 Pro as a structured extraction tool via few-shot prompting.
For each fault type, the extraction process produces a reusable fault-knowledge entry comprising semantic metadata and a parameterized reproduction template.
The metadata records the fault name, causal mechanism, expected local failure state, and possible upper-layer manifestations.
The template specifies abstract prerequisites, the fault-inducing mutation or perturbation, and its activation logic using typed variables rather than workload-specific resource identifiers.
For example, a service selector mismatch template requires the selector of a Service to differ from the labels of its intended backend Pods while leaving the namespace, Service, Pods, label keys, and label values unbound.

Candidate fault-knowledge entries undergo two pre-instantiation checks.
First, an LLM-based consistency check identifies incomplete, ill-formed, or technically inconsistent specifications and refines their structured descriptions.
Second, reviewers with cloud-native operations experience assess the causal logic of each fault and the technical feasibility of its reproduction template.
Runtime validation is then performed on every case-specific instantiation in Phase 2.
A fault type is retained in the released taxonomy only after at least one instantiated case passes this validation.
This separation prevents an extracted template from being treated as executable or valid before it is instantiated and tested in a real cluster.
Ultimately, \benchx contains \faulttype specific root-cause types across 8 categories spanning the Kubernetes stack.

\begin{table*}[t]
\centering
\caption{Fault Taxonomy in \benchx. Easy/Medium/Hard correspond to the diagnostic reasoning complexity.}
\vspace{-0.1in}
\label{tab:root_list}
\resizebox{\textwidth}{!}{
\begin{tabular}{p{6cm} p{14cm} c c} 
\toprule
\textbf{Fault Category} \newline \textit{\footnotesize(Mechanism Description)} & 
\textbf{Specific Fault Types (Full List)} & 
\textbf{Difficulty} & 
\textbf{\# Cases} \\
\midrule

\textbf{Admission Control} \newline
\footnotesize{Requests rejected by API server due to quota or permission violations.} & 
\raggedright 
\href{https://notes.kodekloud.com/docs/Kubernetes-Troubleshooting-for-Application-Developers/Troubleshooting-Scenarios/Case-of-the-Missing-Pods}{NamespaceCPUQuotaExceeded},  \href{https://notes.kodekloud.com/docs/Kubernetes-Troubleshooting-for-Application-Developers/Troubleshooting-Scenarios/Case-of-the-Missing-Pods}{NamespaceMemoryQuotaExceeded},  \href{https://notes.kodekloud.com/docs/Kubernetes-Troubleshooting-for-Application-Developers/Troubleshooting-Scenarios/Case-of-the-Missing-Pods}{NamespacePodQuotaExceeded},  \href{https://notes.kodekloud.com/docs/Kubernetes-Troubleshooting-for-Application-Developers/Troubleshooting-Scenarios/Case-of-the-Missing-Pods}{NamespaceServiceQuotaExceeded}, \href{https://notes.kodekloud.com/docs/Kubernetes-Troubleshooting-for-Application-Developers/Troubleshooting-Scenarios/Case-of-the-Missing-Pods}{NamespaceStorageQuotaExceeded},  \href{https://notes.kodekloud.com/docs/Kubernetes-Troubleshooting-for-Application-Developers/Troubleshooting-Scenarios/Case-of-the-Missing-Pods}{MissingServiceAccount} & 
\textcolor{orange!90!black}{\textbf{Medium}} & 
58 \\
\midrule

\textbf{Scheduling} \newline
\footnotesize{Pods remain Pending due to unsatisfied scheduling constraints or volume-binding requirements.} & 
\raggedright
\href{https://kubernetes.io/docs/reference/kubectl/generated/kubectl_cordon/}{NodeCordonMismatch}, \href{https://notes.kodekloud.com/docs/CKA-Certification-Course-Certified-Kubernetes-Administrator/Scheduling/Node-Affinity}{NodeAffinityMismatch},  \href{https://notes.kodekloud.com/docs/Kubernetes-Troubleshooting-for-Application-Developers/Troubleshooting-Scenarios/Pending-Pods}{NodeSelectorMismatch}, \href{https://kubernetes.io/docs/concepts/scheduling-eviction/assign-pod-node/\#affinity-and-anti-affinity}{PodAntiAffinityConflict},  \href{https://notes.kodekloud.com/docs/Kubernetes-Troubleshooting-for-Application-Developers/Troubleshooting-Scenarios/Pending-Pods}{TaintTolerationMismatch}, \href{https://notes.kodekloud.com/docs/Kubernetes-Troubleshooting-for-Application-Developers/Troubleshooting-Scenarios/Pending-Pods}{CPUCapacityMismatch}, \href{https://notes.kodekloud.com/docs/Kubernetes-Troubleshooting-for-Application-Developers/Troubleshooting-Scenarios/Pending-Pods}{MemoryCapacityMismatch}, \href{https://notes.kodekloud.com/docs/Certified-Kubernetes-Application-Developer-CKAD/State-Persistence/Persistent-Volume-Claims}{PVBindingOccupied}, \href{https://docs.redhat.com/en/documentation/openshift_online/3/html/developer_guide/dev-guide-selector-label-volume-binding}{PVCSelectorMismatch}, \href{https://kubernetes.io/docs/concepts/storage/persistent-volumes/\#reserving-a-persistentvolume}{PVCStorageClassMismatch}, \href{https://medium.com/@veerababu.narni232/kubernetes-volumes-storage-pv-pvc-and-storage-class-548a5ff86343}{PVCCapacityMismatch}, \href{https://notes.kodekloud.com/docs/Certified-Kubernetes-Application-Developer-CKAD/State-Persistence/Solution-Persistent-Volumes-and-Persistent-Volume-Claims-optional}{PVCAccessModeMismatch}  &
\textcolor{green!60!black}{\textbf{Easy}} & 
164 \\
\midrule

\textbf{Startup} \newline
\footnotesize{Containers fail to initialize or start.} & 
\raggedright
\href{https://discuss.kubernetes.io/t/write-permissions-on-volume-mount-with-security-context-fsgroup-option/16524/2}{VolumeMountPermissionDenied},
\href{https://medium.com/@shaikmustafa8991/day-16-100-secret-not-found-ab6cdab8ecd4}{MissingSecretBinding}, \href{https://notes.kodekloud.com/docs/Kubernetes-Troubleshooting-for-Application-Developers/Troubleshooting-Scenarios/Image-Pull-Errors}{IncorrectImageReference},  \href{https://notes.kodekloud.com/docs/Kubernetes-Troubleshooting-for-Application-Developers/Troubleshooting-Scenarios/Image-Pull-Errors}{ImageRegistryDNSFailure},  \href{https://notes.kodekloud.com/docs/Kubernetes-Troubleshooting-for-Application-Developers/Troubleshooting-Scenarios/Create-Container-Errors}{MissingImagePullSecret} & 
\textcolor{green!60!black}{\textbf{Easy}} & 
86 \\
\midrule

\textbf{Runtime} \newline
\footnotesize{Runtime failures involving application execution, health checks, backend access, or replica availability.} & 
\raggedright
\href{https://notes.kodekloud.com/docs/Kubernetes-Troubleshooting-for-Application-Developers/Troubleshooting-Scenarios/Crashing-Pods}{OOMKilled},  \href{https://drdroid.io/stack-diagnosis/kube-probe-tcp-probe-failed--protocol-error}{LivenessProbeIncorrectProtocol},  \href{https://kubernetes.io/docs/tasks/configure-pod-container/configure-liveness-readiness-startup-probes/}{LivenessProbeIncorrectPort},  \href{https://kubernetes.io/docs/tasks/configure-pod-container/configure-liveness-readiness-startup-probes/}{LivenessProbeIncorrectTiming}, \href{https://drdroid.io/stack-diagnosis/kube-probe-tcp-probe-failed--protocol-error}{ReadinessProbeIncorrectProtocol},  \href{https://kubernetes.io/docs/tasks/configure-pod-container/configure-liveness-readiness-startup-probes/}{ReadinessProbeIncorrectPort}, \href{https://sregym.com/problems/sidecar_port_conflict_astronomy_shop}{ServiceSidecarPortConflict}, \href{https://stackoverflow.com/questions/45650044/cant-connect-to-mysql-pod-in-kubernetes-when-using-secrets-for-password-access}{MysqlInvalidCredentials}, \href{https://discuss.kubernetes.io/t/connecting-to-an-external-mysql-database/8201}{MysqlInvalidPort}, \href{https://developer.volcengine.com/articles/7277536281294061629}{DBReadOnly}, \href{https://oneuptime.com/blog/post/2026-01-24-database-connection-microservices/view\#:~:text=Issue\%201:\%20Connection\%20Pool\%20Exhaustion.\%20When\%20all\%20connections\%20are\%20in\%20use}{DBConnectionExhaustion}, \href{https://sregym.com/problems/scale_pod_zero_social_net}{DeploymentZeroReplicas} & 
\textcolor{green!60!black}{\textbf{Easy}} & 
141 \\
\midrule

\textbf{Service Routing} \newline
\footnotesize{Traffic routing failures between internal components.} & 
\raggedright
\href{https://kubernetes.io/docs/tutorials/services/connect-applications-service/}{ServiceSelectorMismatch}, \href{https://support.tools/kubernetes-service-port-mismatches/}{ServicePortMappingMismatch}, \href{https://kubernetes.io/docs/reference/networking/service-protocols/}{ServiceProtocolMismatch}, \href{https://www.plural.sh/blog/kubernetes-env-vars-guide/}{ServiceEnvVarAddressMismatch},\href{https://kubernetes.io/docs/concepts/services-networking/gateway/}{GatewayMisroute},\href{https://kubernetes.io/docs/tasks/administer-cluster/dns-debugging-resolution/}{ServiceDNSResolutionFailure} & 
\textcolor{orange!90!black}{\textbf{Medium}} & 
91 \\
\midrule

\textbf{Performance} \newline
\footnotesize{Performance degradation due to saturation or network drops.} & 
\raggedright
\href{https://www.aiops.cn/gitlab/aiops-live-benchmark/aiopschallenge2025}{PodCPUOverload}, \href{https://www.aiops.cn/gitlab/aiops-live-benchmark/aiopschallenge2025}{PodNetworkDelay},\href{https://www.aiops.cn/gitlab/aiops-live-benchmark/aiopschallenge2025}{NodeNetworkDelay}, \href{https://www.aiops.cn/gitlab/aiops-live-benchmark/aiopschallenge2025}{NodeNetworkPacketLoss} & 
\textcolor{red!70!black}{\textbf{Hard}} & 
76 \\
\midrule

\textbf{Infrastructure} \newline
\footnotesize{Outages in underlying cluster control plane or components.} & 
\raggedright
\href{https://stackoverflow.com/questions/49725980/what-causes-transport-dial-unix-var-run-docker-containerd-docker-containerd-s}{ContainerdUnavailable}, \href{https://github.com/kubernetes/kubernetes/issues/57955}{KubeletUnavailable}, \href{https://stackoverflow.com/questions/61557324/how-to-know-why-the-kube-proxy-stopped}{KubeProxyUnavailable}, \href{https://docs.redhat.com/en/documentation/red_hat_enterprise_linux_atomic_host/7/html/getting_started_with_kubernetes/troubleshooting_kubernetes\#checking_that_kubernetes_systemd_services_are_up}{KubeSchedulerUnavailable} & 
\textcolor{orange!90!black}{\textbf{Medium}} & 
40 \\

\midrule
\textbf{Application Code Defect} \newline
\footnotesize{Application-level code defects in key business logic.} &
\raggedright
\href{https://github.com/IntelligentDDS/ChangeRCA}{CodeMissingParameter}, \href{https://github.com/IntelligentDDS/ChangeRCA}{CodeBusyLoop}, \href{https://github.com/IntelligentDDS/ChangeRCA}{CodeExcessiveFileReads}, \href{https://github.com/IntelligentDDS/ChangeRCA}{CodeWrongReturn}, \href{https://github.com/IntelligentDDS/ChangeRCA}{CodeExcessiveFileWrites}, \href{https://github.com/IntelligentDDS/ChangeRCA}{CodeWrongArgumentOrder}, \href{https://github.com/IntelligentDDS/ChangeRCA}{CodeArtificialDelay}, \href{https://github.com/IntelligentDDS/ChangeRCA}{CodeMemoryLeak} &
\textcolor{red!70!black}{\textbf{Hard}} &
98 \\

\midrule
\textbf{Total} & \textbf{57 distinct fault types} & - & \textbf{754} \\
\bottomrule
\end{tabular}
}
\end{table*}

We further tag each fault category with a diagnostic difficulty label
(Easy/Medium/Hard) that reflects how hard it is to localize the faulty
component and confirm its mechanism from the reported symptom
(Tab.~\ref{tab:root_list}).
The label is assigned at construction time based on how directly the fault is exposed by observable symptoms and how much cross-component or cross-layer investigation is typically needed.
It is assigned at the category level and shared by all cases within that category.

\subsection{Phase 2: Automated Fault Case Generation}\label{sec:phase2}
Traditional fault injection relies heavily on manual orchestration, which is labor-intensive and difficult to scale.
As illustrated in Fig.~\ref{fig:overview}, Phase 2 uses a three-agent closed loop to instantiate the fault-knowledge entries from Phase 1 into executable fault plans and verified benchmark cases.

\textbf{Testbed.}\label{sec:testbed}
To instantiate the fault cases, we deploy a four-node Kubernetes (v1.31) testbed on Huawei Cloud ECS, using \texttt{c6.2xlarge.2} instances with 8 vCPUs and 16~GiB of memory per node.
As the foundation of cloud-native infrastructure, Kubernetes offers not only a complex distributed control plane for fault injection but also a standardized corpus of documentation.
This allows us to rigorously test knowledge-grounded reasoning.
The observability plane integrates Prometheus~\cite{prome} and Istio~\cite{istio} for metrics, alongside native APIs for object states and logs.
As target workloads, we deploy two widely adopted microservice benchmarks, \textbf{OnlineBoutique} (11 microservices)~\cite{OnlineBoutique} and \textbf{TrainTicket} (41 microservices with database-backed services)~\cite{Trainticket}, and generate user traffic with Locust~\cite{locust}.
We use ChaosBlade~\cite{Chaosblade} to inject CPU load, network delay, and packet loss, and \texttt{kubectl} commands to introduce configuration faults.
Together, these mechanisms realize faults across application services and Kubernetes platform layers.
For code-defect faults, we mutate service implementations with representative backend bug patterns, such as missing parameter values, wrong return values, or incorrect exception handling following \cite{changerca2024fse}.
Each mutated implementation is packaged into a service image and deployed to the testbed.

\subsubsection{Fault Injection Loop}
\label{sec:fault_injection_loop}

For each candidate case $j$, the \textit{Generator Agent}, implemented with Gemini 3 Pro, produces a case-specific fault plan $\mathcal{B}_j$ from a fault-knowledge entry and the target workload.
Each plan specifies the required pre-injection state, resource selectors and operational parameters, ordered injection operations, and runtime validation assertions.
The validation assertions cover mechanism activation, the expected local failure state, and any applicable upper-layer manifestation.
Before each injection attempt, the \textit{Executor Agent} restores the testbed to a clean deployment, resolves the selectors against the restored cluster state, establishes the required prerequisites, and confirms the pre-injection health baseline, including a successful end-to-end request for service-facing cases, before applying the current fault plan.

The \textit{Verifier Agent} evaluates the validation assertions against the resulting cluster state using object states, events, logs, metrics, and connectivity results.
It accepts an execution only when the fault-inducing condition is active at the target component, the expected local failure state is present, and any specified upper-layer manifestation is observed.
After successful verification, the resolved target and validated fault condition are recorded in the verified fault plan.
This verification is necessary because Kubernetes recovery mechanisms may mask an injected fault.
For example, tainting one node may fail to induce a scheduling failure because a newly created Pod can still be placed on another eligible node.
If validation fails, the Verifier compares the observed runtime state with the intended fault mechanism and returns structured feedback to the Generator, such as missing prerequisites, insufficient perturbation intensity, or an unintended recovery path.
The Generator may then revise resource selectors or operational parameters while preserving the fault mechanism and validation criteria defined by the fault-knowledge entry.
The clean deployment, injection, and verification procedure then repeats.
Cases that still fail validation after at most three correction attempts are rejected rather than included in the benchmark.

  \begin{table*}[t]
  \centering
  \caption{Diagnostic tools exposed by Cloud-OpsBench.}
  \label{tab:tool_description}
  \vspace{-0.15in}
  \resizebox{1\textwidth}{!}{
      \begin{tabular}{p{8cm} p{10cm}}
      \toprule
      \textbf{Tool Name \footnotesize{(Arguments)}} & \textbf{Returned Evidence} \\
      \midrule
      \texttt{T1.GetResources}\footnotesize{(type, namespace, name, flags)}
      & Resource lists or selected objects with status, metadata, and extended attributes. \\
      \texttt{T2.DescribeResource}\footnotesize{(type, name, namespace)}
      & Runtime details of a Kubernetes object, including state, conditions, and events. \\
      \texttt{T3.GetAppYAML}\footnotesize{(app\_name)}
      & Deployment configuration YAML for the target application service. \\
      \texttt{T4.GetServiceDependencies}\footnotesize{(service\_name)}
      & Service dependency graph rooted at the target service. \\
      \texttt{T5.GetRecentLogs}\footnotesize{(service\_name, namespace)}
      & Recent service logs for runtime error inspection. \\
      \texttt{T6.CheckServiceConnectivity}\footnotesize{(namespace, svc\_name, port)}
      & TCP reachability result for the target service endpoint. \\
      \texttt{T7.GetClusterConfiguration}\footnotesize{()}
      & Cluster-wide node information, including resources, labels, taints, and status. \\
      \texttt{T8.GetAlerts}\footnotesize{()}
      & Metric anomaly alerts generated by threshold-based detectors. \\
      \texttt{T9.GetErrorLogs}\footnotesize{(service\_name, namespace)}
      & Summary of log lines matching error, exception, or failure patterns. \\
      \texttt{T10.CheckNodeServiceStatus}\footnotesize{(node\_name, component)}
      & Liveness status of node-level or control-plane components. \\
      \texttt{T11.ListCodeFiles}\footnotesize{(app\_name)}
      & Available source files for an application service with brief file descriptions. \\
      \texttt{T12.GetSourceCode}\footnotesize{(app\_name, file\_path)}
      & Source code content for a selected application file. \\
      \bottomrule
      \end{tabular}
  }
  \vspace{-0.2in}
  \end{table*}

A case is admitted to the snapshot-construction stage only after all runtime validation assertions are satisfied.
After the Verifier confirms the observable manifestation used to define the task, we formulate $U_j$ from that manifestation while excluding the faulty component, fault mechanism, and other construction-only information.
The verified fault plan is retained as construction provenance but is neither exposed to the evaluated agent nor accepted as diagnostic evidence.
Across the construction process, 800 candidate fault plans were generated, and 46 were rejected because they failed to manifest stable or diagnosable symptoms.


\subsubsection{State Snapshots}

After a fault case passes runtime verification, \benchx materializes its tool-observable context as a state snapshot (\S~\ref{sec:design_and_formulation}).
The snapshot captures system state after verification and retains timestamped events, logs, and metrics from the corresponding case-specific incident interval.
The snapshot builder records the baseline-confirmation and manifestation-verification timestamps and applies these boundaries when retrieving temporal evidence.
To construct the replay interface, we pre-execute the diagnostic toolset (\texttt{T1}--\texttt{T12}, Table~\ref{tab:tool_description}) against this verified incident context and store the resulting tool responses in a JSON repository.
This repository serves as the persistence layer for deterministic replay.
The snapshot-backed replay is therefore not a hand-written simulator.
During evaluation, it returns the corresponding precomputed response.
Before lookup, the interface validates each request against the supported tool schema and canonicalizes its critical arguments so that equivalent requests map to the same repository entry.
Replay coverage is limited to the tool queries supported by the preserved snapshot.
Queries outside this coverage receive tool-specific error or empty-result responses.

The snapshot builder covers four evidence channels.
First, resource state and events are captured by executing retrieval and description tools (\texttt{T1}, \texttt{T2}) over Kubernetes objects in the target namespace.
Second, application topology and telemetry are captured by retrieving static configurations (\texttt{T3}), dependency graphs (\texttt{T4}), logs (\texttt{T5}, \texttt{T9}), and reachability results (\texttt{T6}).
Third, cluster-level context is captured by collecting node and component health information (\texttt{T7}, \texttt{T10}) and converting metrics into structured alerts (\texttt{T8}).
Fourth, source-code evidence is captured by listing available code files and retrieving source code (\texttt{T11}, \texttt{T12}).
These four channels cover the standard evidence sources used in Kubernetes troubleshooting.

Although replay responses remain fixed during evaluation, timestamped logs, Kubernetes events, and alerts comparing normal and abnormal operation preserve temporal information about the incident.
For faults such as memory leaks and database connection exhaustion, agents can combine this information with resource states, configurations, and, where available, source code to identify the underlying mechanism.

This design provides faithful replay with an explicit coverage boundary.
For supported queries, the interface returns outputs captured from a verified incident in a real cluster rather than responses generated by a manually encoded model of Kubernetes behavior.
Coverage is finite and measurable because the construction sweep enumerates the supported tools over case-valid arguments derived from the objects, namespaces, service endpoints, nodes, and source files available in each verified case.
Snapshots for OnlineBoutique and TrainTicket contain an average of 599 and 1,708 precomputed responses per case and occupy 3.46 and 8.75~MiB, respectively.
A case is retained only when subsequent annotation validation confirms that all required diagnostic facts are observable through the snapshot-backed interface.
The finite replay repository bounds the observable evidence, but agents must independently choose tools and arguments as the investigation unfolds rather than follow a prescribed query list or command sequence.
Through this construction, state snapshots address Gap~1 and Gap~2: agents investigate via system-facing diagnostic tools rather than offline telemetry dumps, while different agents face the same runtime-verified incident evidence and deterministic query--response mapping.

\vspace{-0.1cm}

\subsection{Phase 3: Ground-Truth Annotation}
\label{sec:gt_graph}
Given the verified fault cases and snapshots produced in Phase 2, we annotate both the root cause and the diagnostic evidence needed to support it.
For each verified case $j$, \benchx defines outcome and process labels,
\[
\mathcal{G}_j^*
=
\left\langle
\mathcal{R}_j^*,
\mathcal{D}_j^*
\right\rangle .
\]
The outcome label
$\mathcal{R}_j^*=\langle C_j^*,F_j^*\rangle$
identifies the faulty component and fault type and is derived from the runtime-verified fault plan $\mathcal{B}_j$.
The process label $\mathcal{D}_j^*$ specifies the diagnostic evidence process associated with $\mathcal{R}_j^*$.

\subsubsection{Process Representation.}
For each case $j$, the process ground truth is represented as
\[
\mathcal{D}_j^*
=
\left\langle
\mathcal{M}_j^*,
\phi_j^*,
\prec_j^*,
\mathcal{Q}_j^*
\right\rangle ,
\]
where $\mathcal{M}_j^*$ denotes the set of diagnostic milestones, $\phi_j^*$ specifies the completion logic over these milestones, $\prec_j^*$ defines prerequisite relations among milestones, and $\mathcal{Q}_j^*$ specifies the admissible evidence bundles for establishing each milestone.

Each milestone $m\in\mathcal{M}_j^*$ represents a diagnostic fact to be established during the investigation. The case-specific completion formula $\phi_j^*$ combines milestones using conjunction and disjunction. Conjunctive terms specify milestones or branches that must be completed jointly, whereas disjunctive terms specify alternative completion paths. The diagnostic evidence process is complete when $\phi_j^*$ evaluates to true over the set of established milestones.
The directed acyclic relation $\prec_j^*$ records immediate prerequisite relations among milestones. For $m_a\prec_j^*m_b$, milestone $m_a$ is a prerequisite of $m_b$. Its transitive closure defines the corresponding partial order, while milestones not connected by a dependency path remain unordered.

For example, consider $\phi_j^*=M_1\land[M_2\lor(M_3\land M_4)]$, with prerequisite relations $M_1\prec_j^*M_2$, $M_1\prec_j^*M_3$, and $M_3\prec_j^*M_4$. Here, $\phi_j^*$ defines the milestone combinations that satisfy the completion condition, while $\prec_j^*$ specifies the prerequisite structure among these milestones.

For each milestone $m$, $\mathcal{Q}_j^*(m)$ contains the admissible evidence bundles that can establish it. Each atomic evidence clause $q$ specifies a diagnostic tool, admissible argument configurations, and, where applicable, a predicate over the normalized response. An evidence bundle is satisfied when all of its clauses have been observed, and a milestone is established when at least one admissible bundle is satisfied.
\begin{table}[t]
\centering
\caption{Representative mechanism-relevant objects and states by fault category.}
\label{tab:evidence_scope}
\small
\setlength{\tabcolsep}{5pt}
\renewcommand{\arraystretch}{1.12}
\vspace{-0.15in}
\resizebox{0.8\textwidth}{!}{%
\begin{tabular}{@{}>{\raggedright\arraybackslash}p{0.25\linewidth}>{\raggedright\arraybackslash}p{\dimexpr0.75\linewidth-2\tabcolsep\relax}@{}}
\toprule
\textbf{Fault Category} & \textbf{Relevant Objects / States} \\
\midrule
\textbf{Admission Control} & Namespace quotas, resource requests, service-accounts, resource creation states \\
\addlinespace[2pt]
\textbf{Scheduling} & Pod scheduling states/events, workload--node constraints, PV/PVC requirements and binding states \\
\addlinespace[2pt]
\textbf{Startup} & Container startup states, failure logs, image-pull configuration, Secret references, volume/mount configuration \\
\addlinespace[2pt]
\textbf{Runtime} & Workload replica, container lifecycle states, resource limits, probes, application/database configuration, runtime error logs \\
\addlinespace[2pt]
\textbf{Service Routing} & Service configuration, gateway routes, application connection addresses, DNS configuration, backend Pods, endpoint states, service connectivity \\
\addlinespace[2pt]
\textbf{Performance} & Resource/performance alerts, Pod/node health and placement, service dependencies \\
\addlinespace[2pt]
\textbf{Infrastructure} & Node health, Kubernetes component states, container runtime states, Pod states, service connectivity \\
\addlinespace[2pt]
\textbf{Application Code Defect} & Source code, container health, runtime error logs, performance symptoms, service dependencies \\
\bottomrule
\end{tabular}
}
\vspace{-0.15in}
\end{table}

\subsubsection{Annotation Construction.}
Our process annotations target a complete evidence chain from the reported symptom to the underlying fault mechanism.
The milestones capture relevant symptoms, component relationships, and faulty configuration or runtime evidence that explain how the fault produced the observed impact.
This evidence chain provides a basis for diagnostic reporting and subsequent remediation.
Annotations are constructed in three stages.

\textit{Step 1: Constrain the candidate evidence space.}
Each verified case is associated with a known fault mechanism, which determines the system objects and runtime states relevant to its diagnosis. These case-relevant states are preserved in the frozen snapshot and together form a finite candidate evidence space from which diagnostic milestones and their supporting evidence are identified.
Table~\ref{tab:evidence_scope} summarizes the principal object and state families associated with the eight fault categories.

\textit{Step 2: Identify and structure diagnostic milestones.}
Annotators identify diagnostic facts from the candidate evidence space and encode them as milestones in $\mathcal{M}_j^*$.
Facts that can be established independently are represented as separate milestones.
The milestones are then organized by the completion formula $\phi_j^*$.
Conjunctive branches encode evidence that must be established jointly, while disjunctive branches encode alternative diagnostic paths.
Prerequisite relations among milestones are recorded in $\prec_j^*$, producing the partial-order structure used for process evaluation.

\textit{Step 3: Enumerate admissible evidence for each milestone.}
For each milestone, because each snapshot repository is finite and exhaustively indexed, we systematically scan all captured responses across T1–T12 to identify every tool output that exposes the target diagnostic fact. Each evidence clause specifies a diagnostic tool, one or more valid argument configurations, and, when necessary, the diagnostic observation that must be present in the returned result. A milestone may therefore be established through different tool choices, query scopes, or argument configurations that expose the same diagnostic evidence. When a single observation is sufficient, it forms a singleton evidence bundle; when multiple complementary observations are required, they are grouped into a joint bundle. All admissible evidence bundles for milestone $m$ are recorded in $\mathcal{Q}_j^*(m)$.

\begin{figure*}[t]
    \centering
    \includegraphics[width=\textwidth]{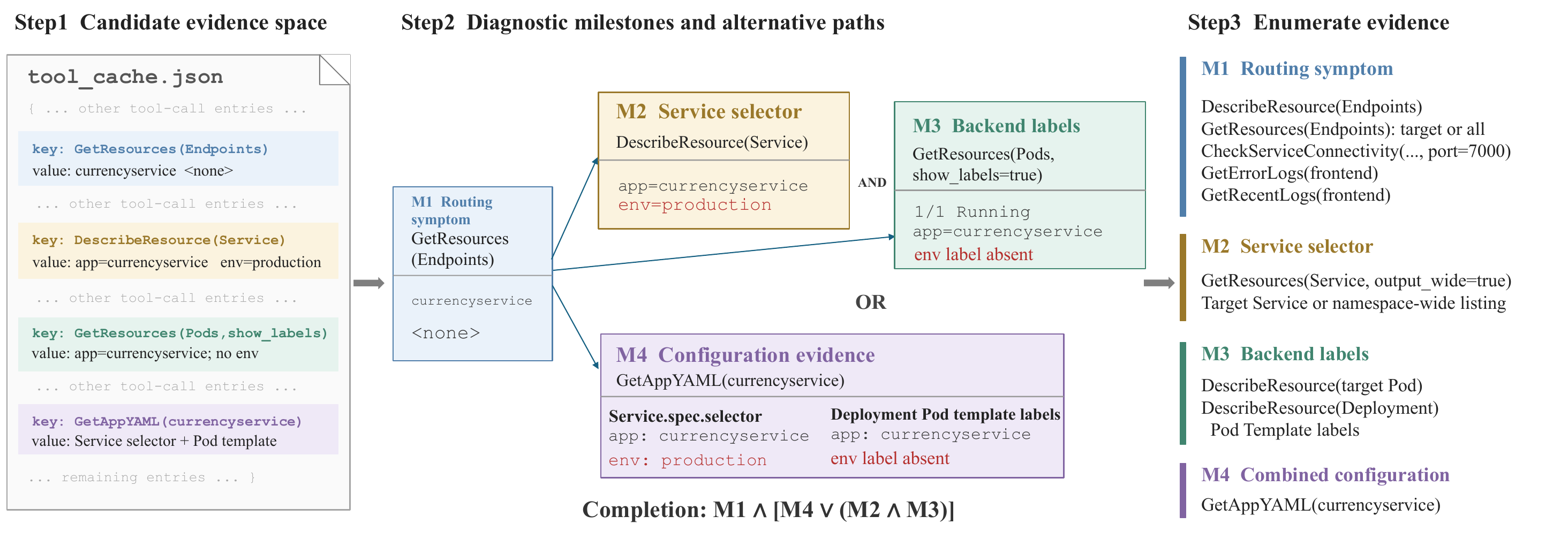}
    \vspace{-0.1in}
    \caption{Annotation construction for OnlineBoutique case \texttt{service/50}: identifying relevant observations in the frozen snapshot, structuring diagnostic milestones and alternative completion paths, and enumerating admissible evidence.}
    \label{fig:annotation}
\end{figure*}

Figure~\ref{fig:annotation} illustrates the three annotation stages using OnlineBoutique case \texttt{service/50}, a Service selector mismatch affecting \texttt{currencyservice}.
The known fault mechanism directs annotation to the Service, its endpoints, and the backend workload preserved in the snapshot.
The relevant observations establish a routing symptom, illustrated by the empty endpoint set ($M_1$), a Service selector requiring both \texttt{app=currencyservice} and \texttt{env=production} ($M_2$), and backend Pod labels lacking \texttt{env=production} ($M_3$).
The application YAML provides an alternative source of combined configuration evidence ($M_4$), exposing both the Service selector and the Deployment's Pod-template labels.
The completion formula is therefore $M_1 \land [M_4 \lor (M_2 \land M_3)]$.
The annotated dependencies place $M_1$ before the configuration investigation, while leaving $M_2$ and $M_3$ unordered.

The admissible evidence bundles connect these milestones to concrete tool calls and their returned observations.
For example, $M_2$ can be established by describing the Service or retrieving its wide resource listing.
For $M_3$, the backend labels can be obtained through a Pod listing with \texttt{show\_labels=true}, a description of the target Pod, or the Pod-template information in the Deployment description.
These alternatives establish the same diagnostic facts through different queries.

\textbf{Annotation Validation.}
Each annotation is first automatically checked against its frozen snapshot. The validation verifies the structural consistency of the completion formula and prerequisite relations, and replays the admissible evidence bundles to ensure that their tool calls and returned observations are valid for the corresponding case.
Two auditors with practical experience in Kubernetes operations and cloud fault diagnosis independently review each annotation for diagnostic sufficiency, milestone granularity, completion-logic validity, and prerequisite correctness, specifically verifying whether alternative tool channels (e.g., startup logs versus object descriptions or YAML manifests) could independently establish the same milestone. They inspect $\mathcal{R}_j^*$, $\mathcal{D}_j^*$, and the corresponding frozen snapshot. Disagreements and proposed revisions are adjudicated by a third senior practitioner with expertise in cloud-native operations.

Following independent audit and adjudication, 9.4\% of the annotations were revised and automatically revalidated. The retained annotations contain, on average, 3.0 milestones and 3.1 admissible evidence bundles per milestone.

\section{Evaluation}
In this section, we evaluate LLM agents on \benchx to answer the following RQs:
\begin{enumerate}[leftmargin=*,label=\textbullet]
  \item \textbf{RQ1 (Outcome Effectiveness):} How accurately do agents identify the faulty component and fault type?
  \item \textbf{RQ2 (Evidence Process):} Do agents collect the required evidence and follow the annotated diagnostic prerequisites?
  \item \textbf{RQ3 (Experience Reuse):} Can RCA agents benefit from historical diagnostic experience?
  \item \textbf{RQ4 (Agent Architectures and Harnesses):} How do different agent implementations affect diagnostic accuracy, evidence completeness, and computational cost?
\end{enumerate}
Additionally, we analyze recurring failure modes in incorrect diagnoses and evidence gaps in correct diagnoses through trajectory analysis.


\subsection{Experimental Setup}


\textbf{Setup and Models.} We implement a baseline agent framework based on the standard ReAct~\cite{yao2022react} paradigm, utilizing Pydantic~\cite{pydantic} for strict tool parameter validation. This custom architecture natively ensures full-trace observability of the agent's reasoning trajectories and tool invocations. For the ReAct-based experiments in RQ1--RQ3, we use a maximum of 20 interaction steps and set the temperature to 0 to reduce sampling variability.
The agent configurations and diagnostic tool-call budget used in RQ4 are specified separately.
We evaluate eleven models covering both \textbf{proprietary LLMs} and \textbf{open-weight LLMs}.
The proprietary LLMs include GPT-5, Claude-Sonnet-4, Gemini-2.5-Pro, Gemini-2.5-Flash, and Qwen3.5-Plus.
The open-weight LLMs include GLM-5.1, DeepSeek-V4-Flash, Qwen3-235B-A22B, Qwen3.5-27B, Qwen3-14B, and Qwen3-8B, covering high-capacity and compact variants\cite{SLM}.

\textbf{Diagnostic-difficulty strata.}
We report selected results stratified by the construction-time diagnostic
difficulty labels defined in \S~\ref{sec:phase1}
(Tab.~\ref{tab:root_list}).
Each evaluation case inherits its category label; we do not reassign
difficulty from agent scores.
We report each model's results separately within each system and difficulty group.

\subsubsection{Evaluation Metrics}We evaluate agents along three complementary dimensions: outcome correctness, process grounding, and diagnostic efficiency.

\textbf{(1) Outcome Correctness.}
Our primary metric is \textbf{Joint RCA Accuracy} (JRA), the fraction of episodes for which $\hat{\mathcal{R}}_j=\mathcal{R}_j^*$, equivalently, $\hat{C}_j=C_j^*$ and $\hat{F}_j=F_j^*$.
We also report \textbf{Component Accuracy} (CA) and \textbf{Fault-Type Accuracy} (FA) to distinguish localization errors from fault-type classification errors.
An episode with a missing or invalid field is counted as incorrect for the corresponding accuracy metric.

\textbf{(2) Process Grounding.}
We evaluate each trajectory $\mathcal{T}_j$ against its diagnostic evidence graph $\mathcal{D}_j^*$ (\S~\ref{sec:gt_graph}), following process supervision~\cite{prm} and milestone-based agent evaluation~\cite{WebCanvas,AgentBoard}.
A milestone is established when the tool observations satisfy at least one of its admissible evidence bundles.
Let $\mathcal{M}_j^{\mathrm{est}}$ denote the set of established milestones.

\textbf{Milestone Coverage} (MC) measures partial completion of the annotated completion formula $\phi_j^*$.
An established milestone receives a score of 1, and an unestablished milestone receives 0.
We evaluate the formula from its milestones upward: each AND node takes the mean score of its child requirements, and each OR node takes the highest score among its alternative branches.
Let $s(\phi,S)$ denote the resulting score for formula $\phi$ when the established milestone set is $S$.
We define $\mathrm{MC}_j=s(\phi_j^*,\mathcal{M}_j^{\mathrm{est}})$.
MC therefore credits alternative diagnostic paths without requiring every branch to be completed, and additional exploratory calls do not reduce its value.

\textbf{Evidence-Order Consistency} (EOC) measures milestone coverage under the annotated diagnostic prerequisites.
We scan the trajectory chronologically.
A milestone receives order credit when its evidence bundle is completed or refreshed after, or in the same tool call as, all its prerequisites receive order credit.
Evidence collected too early can receive credit through a later observation that completes or refreshes its bundle after the prerequisites are established.
Let $\mathcal{M}_j^{\mathrm{ord}}$ contain the milestones satisfying this rule.
We compute $\mathrm{EOC}_j=s(\phi_j^*,\mathcal{M}_j^{\mathrm{ord}})$ using the same scoring function as MC.
Independent milestones may be investigated in either order.

\textbf{Evidence Closure Rate} (ECR) measures whether an investigation completes the annotated evidence requirements.
For each case, ECR equals 1 when the completion formula is fully satisfied, equivalently when $\mathrm{MC}_j=1$, and 0 otherwise.
Completing one alternative path is sufficient; evidence acquisition order is evaluated separately by EOC.
We report MC and EOC as means over evaluated cases, and ECR as the proportion of cases satisfying the closure condition.

\textbf{(3) Diagnostic Efficiency.}
\textbf{Evidence Efficiency} (EE) measures the proportion of diagnostic tool calls that establish or refresh milestone evidence.
A call receives credit when its observation satisfies an admissible singleton evidence bundle or completes or refreshes a joint evidence bundle.
Each call is counted at most once, even if it matches multiple milestones.
Repeated calls may receive EE credit when they satisfy the same evidence conditions; acquisition order is evaluated separately by EOC.

\textbf{Redundant Action Rate} (RAR) measures the proportion of diagnostic tool calls that repeat a previously issued tool-and-argument pair.
It complements EE by identifying repeated queries, including those that retrieve milestone-relevant evidence.

We also report \textbf{Steps} to measure investigation length.
In the main ReAct experiments, this count includes diagnostic tool calls and the final diagnosis submission.
In RQ4, Steps counts diagnostic tool calls only, excluding planning and control calls.
To compare the cost-efficiency of different agent architectures and harnesses, we additionally report \textbf{Tokens}, the mean total input and output tokens consumed by all model calls per case, expressed in thousands (K).

\begin{figure*}[h]
    \centering
    \includegraphics[width=0.85\textwidth]{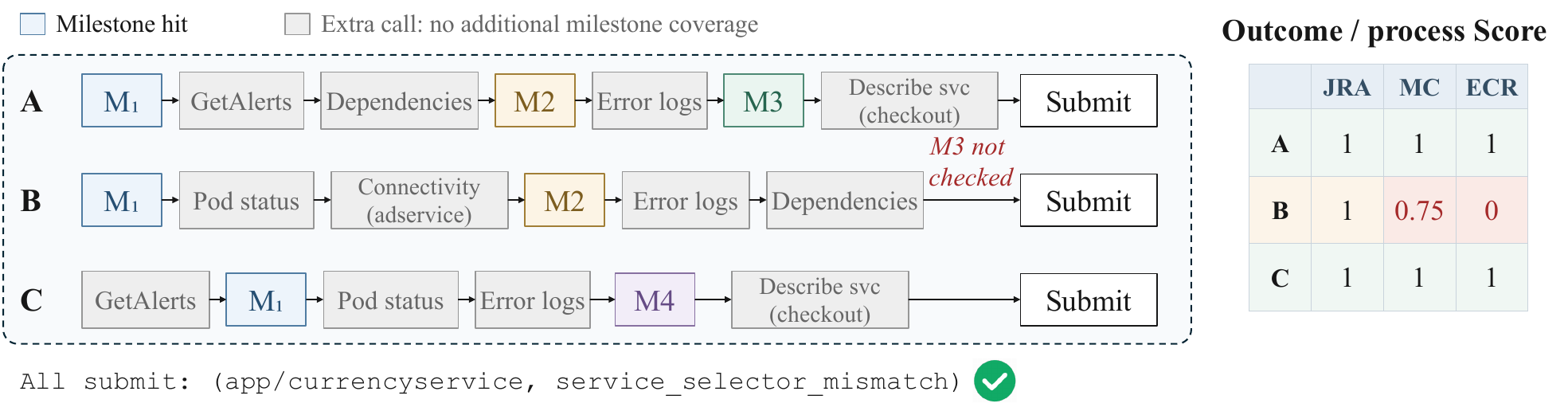}
    \vspace{-0.1in}
    \caption{Illustrative trajectories with the same correct RCA but different evidence coverage. A and C complete alternative evidence paths, whereas B omits backend-label evidence; gray steps indicate additional exploration that does not change MC or ECR.}
    \label{fig:case-score}
\end{figure*}

Figure~\ref{fig:case-score} illustrates the scoring rules for the case annotated in Figure~\ref{fig:annotation}, with all three trajectories submitting the correct diagnosis ($\text{JRA}=1$). Trajectory A establishes $M_1$, $M_2$, and $M_3$, completing the selector--label comparison to achieve $\text{MC}=1$ and $\text{ECR}=1$. Trajectory C reaches the same $\text{MC}=1$ and $\text{ECR}=1$ via the alternative configuration branch ($M_1$ and $M_4$). In contrast, Trajectory B establishes only $M_1$ and $M_2$ while omitting the backend-label check $M_3$; under $\phi_j^* = M_1 \wedge [M_4 \vee (M_2 \wedge M_3)]$, it receives partial coverage $\text{MC}_B = [1 + \max(0, (1+0)/2)]/2 = 0.75$ and $\text{ECR}=0$. Across all trajectories, the gray steps represent extra exploratory calls that increase investigation length (Steps) and reduce efficiency (EE) without penalizing MC or ECR.

\subsection{RQ1: RCA Outcome Effectiveness}
\label{sec:rq1_analysis}

Table~\ref{tab:main} compares eleven models under the shared ReAct scaffold and the 20-step budget.
GLM-5.1 achieves the highest JRA on OnlineBoutique (0.78), while GPT-5 leads on TrainTicket (0.68).
The relative performance of individual models also changes across workloads: GLM-5.1 reaches 0.53 JRA on TrainTicket, whereas GPT-5 obtains 0.68 on both systems.
Open-weight models are competitive in this setting.
DeepSeek-V4-Flash reaches 0.73 JRA on OnlineBoutique and 0.63 on TrainTicket.
Qwen3.5-27B reaches 0.72 JRA on OnlineBoutique, compared with 0.68 for GPT-5 and 0.67 for Gemini-2.5-Pro.
These results provide useful reference points for model selection; the component, fault-type, and difficulty breakdowns below show which aspects of diagnosis contribute to the aggregate scores.

\textbf{(1) Locating the faulty component and identifying its fault mechanism present different challenges.}
On OnlineBoutique, every evaluated model achieves higher CA than FA.
GPT-5 obtains 0.86 CA, 0.69 FA, and 0.68 JRA, while GLM-5.1 obtains 0.88 CA, 0.79 FA, and 0.78 JRA.
For these models, approximately 18\% and 10\% of cases, respectively, have a correct component prediction but an incorrect or missing fault type.
Agents can therefore locate the faulty component without correctly explaining what has gone wrong.
Pod states, events, resource configurations, and application logs often identify an affected component directly.
However, different faults in the same component can produce similar symptoms.
Identifying the specific mechanism requires further inspection of the relevant configuration and runtime evidence.

TrainTicket also exposes errors in the opposite direction.
Qwen3.5-Plus achieves 0.79 FA but 0.67 CA and 0.64 JRA.
Approximately 15\% of cases therefore have a correct fault type but an incorrect or missing component.
TrainTicket includes gateways and shared database dependencies, allowing a fault in one component to produce errors in several others.
For example, a database failure may appear in an application service's logs, requiring the agent to distinguish the service reporting the error from the dependency responsible for it.
Recognizing the failure mechanism is insufficient when its source is attributed to the wrong component.
JRA requires both parts of the diagnosis to be correct.

\begin{table*}[t]
\centering
\caption{Performance comparison of LLM agents on outcome and process metrics.The best results in each metric are bold.}
\vspace{-0.1in}
\label{tab:main}
\scriptsize
\setlength{\tabcolsep}{2pt}
\resizebox{\textwidth}{!}{
\begin{tabular}{lccccccccc|ccccccccc}
\toprule
& \multicolumn{9}{c}{\cellcolor[HTML]{DAE8FC}\textbf{OnlineBoutique}} & \multicolumn{9}{c}{\cellcolor[HTML]{FFFDD9}\textbf{TrainTicket}} \\
\cmidrule(lr){2-10}\cmidrule(lr){11-19}
& \multicolumn{3}{c}{\textbf{Outcome}} & \multicolumn{6}{c}{\textbf{Process}} & \multicolumn{3}{c}{\textbf{Outcome}} & \multicolumn{6}{c}{\textbf{Process}} \\
\cmidrule(lr){2-4}\cmidrule(lr){5-10}\cmidrule(lr){11-13}\cmidrule(lr){14-19}
\textbf{Model} &
CA$\uparrow$ & FA$\uparrow$ & JRA$\uparrow$ & MC$\uparrow$ & EOC$\uparrow$ & ECR$\uparrow$ & EE$\uparrow$ & Steps$\downarrow$ & RAR$\downarrow$ &
CA$\uparrow$ & FA$\uparrow$ & JRA$\uparrow$ & MC$\uparrow$ & EOC$\uparrow$ & ECR$\uparrow$ & EE$\uparrow$ & Steps$\downarrow$ & RAR$\downarrow$ \\
\midrule
GPT-5 & 0.86 & 0.69 & 0.68 & 0.70 & 0.62 & 0.39 & \textbf{0.64} & \textbf{5.56} & \textbf{0.00} & \textbf{0.70} & 0.71 & \textbf{0.68} & 0.72 & 0.68 & 0.40 & \textbf{0.64} & 7.03 & \textbf{0.00} \\
GLM-5.1 & \textbf{0.88} & \textbf{0.79} & \textbf{0.78} & \textbf{0.84} & 0.74 & \textbf{0.65} & 0.54 & 7.75 & \textbf{0.00} & 0.65 & 0.58 & 0.53 & 0.65 & 0.63 & 0.35 & 0.38 & 8.83 & 0.02 \\
Claude-Sonnet-4 & 0.82 & 0.67 & 0.66 & 0.82 & 0.75 & 0.64 & 0.41 & 9.47 & 0.01 & 0.63 & 0.64 & 0.62 & \textbf{0.76} & 0.71 & 0.15 & 0.52 & 7.42 & \textbf{0.00} \\
Gemini-2.5-Pro & 0.81 & 0.69 & 0.67 & 0.79 & 0.74 & 0.62 & 0.61 & 6.21 & \textbf{0.00} & 0.63 & 0.61 & 0.58 & 0.73 & 0.67 & 0.40 & 0.55 & 7.13 & \textbf{0.00} \\
Qwen3.5-Plus & 0.74 & 0.64 & 0.59 & 0.72 & 0.67 & 0.44 & 0.59 & 6.39 & \textbf{0.00} & 0.67 & \textbf{0.79} & 0.64 & 0.73 & 0.69 & 0.41 & 0.63 & 5.90 & \textbf{0.00} \\
Qwen3-235B-A22B & 0.62 & 0.45 & 0.39 & 0.55 & 0.47 & 0.34 & 0.33 & 7.22 & 0.02 & 0.47 & 0.27 & 0.26 & 0.62 & 0.60 & 0.33 & 0.34 & 8.67 & 0.03 \\
DeepSeek-V4-Flash & 0.86 & 0.74 & 0.73 & 0.80 & 0.71 & 0.57 & 0.58 & 7.24 & \textbf{0.00} & 0.69 & 0.71 & 0.63 & \textbf{0.76} & \textbf{0.72} & \textbf{0.51} & 0.57 & 7.74 & \textbf{0.00} \\
Gemini-2.5-Flash & 0.78 & 0.64 & 0.63 & 0.77 & 0.74 & 0.56 & 0.55 & 6.29 & 0.01 & 0.52 & 0.46 & 0.42 & 0.51 & 0.46 & 0.23 & 0.46 & \textbf{4.95} & \textbf{0.00} \\
Qwen3.5-27B & 0.86 & 0.73 & 0.72 & 0.82 & \textbf{0.78} & 0.64 & 0.56 & 6.97 & 0.01 & 0.63 & 0.61 & 0.56 & 0.71 & 0.68 & 0.42 & 0.57 & 7.03 & 0.01 \\
Qwen3-14B & 0.61 & 0.43 & 0.42 & 0.72 & 0.66 & 0.55 & 0.32 & 14.18 & 0.26 & 0.42 & 0.25 & 0.25 & 0.66 & 0.61 & 0.39 & 0.32 & 14.86 & 0.31 \\
Qwen3-8B & 0.38 & 0.22 & 0.21 & 0.57 & 0.45 & 0.19 & 0.34 & 15.78 & 0.45 & 0.43 & 0.30 & 0.26 & 0.62 & 0.61 & 0.26 & 0.40 & 11.50 & 0.22 \\
\bottomrule
\end{tabular}
}
\end{table*}

\begin{figure*}[t]
    \centering
    \includegraphics[width=1\textwidth]{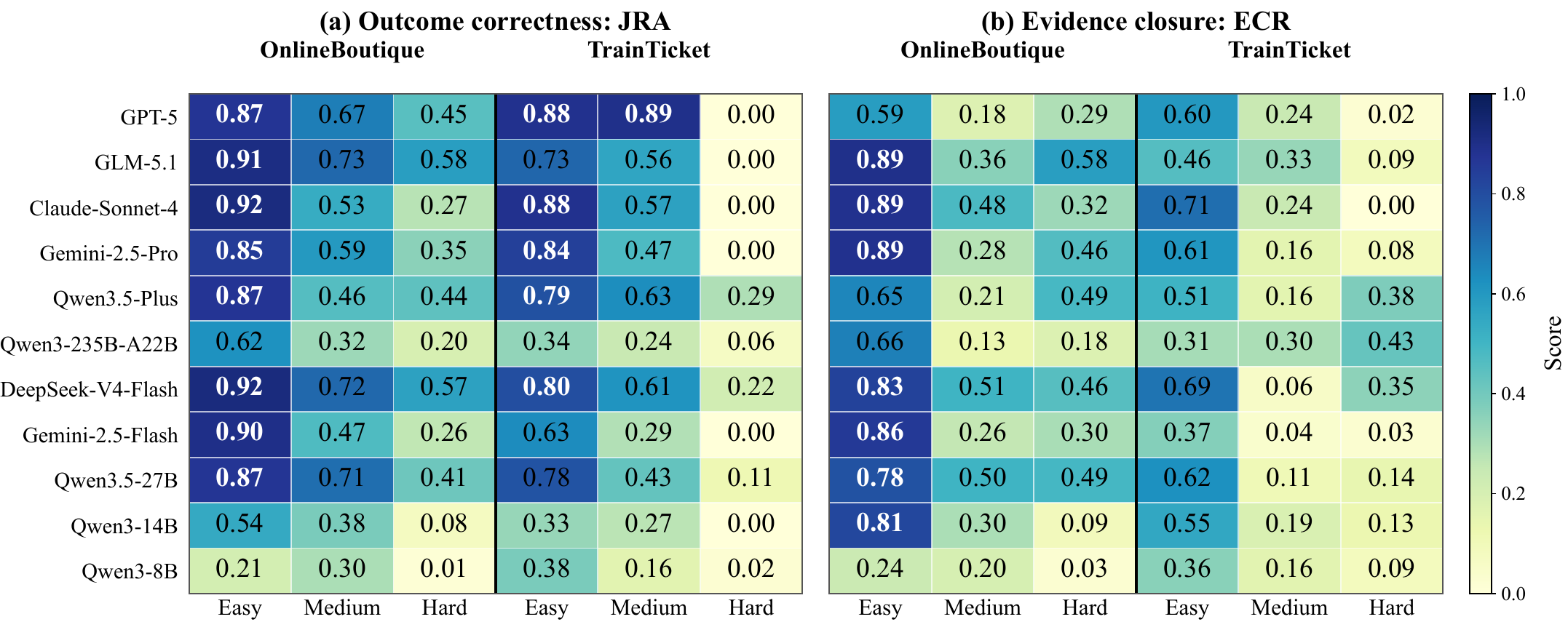}
    \vspace{-0.1in}
    \caption{JRA (a) and ECR (b) of LLM agents on OnlineBoutique and TrainTicket across Easy, Medium, and Hard faults.}
    \label{fig:difficulty}
\end{figure*}

\textbf{(2) Diagnostic accuracy declines across the predefined difficulty strata.}
Figure~\ref{fig:difficulty}(a) presents JRA across the difficulty groups defined in Table~\ref{tab:root_list}.
Several models achieve high accuracy on Easy faults, but their performance generally declines on the more difficult groups.
On OnlineBoutique, GLM-5.1 achieves 0.91, 0.73, and 0.58 JRA across Easy, Medium, and Hard faults.
DeepSeek-V4-Flash obtains 0.92, 0.72, and 0.57, while Qwen3.5-27B obtains 0.87, 0.71, and 0.41.

The difficulty increase is particularly pronounced on TrainTicket.
GPT-5 achieves 0.88 JRA on Easy faults and 0.89 on Medium faults, but 0.00 on Hard faults.
GLM-5.1 obtains 0.73, 0.56, and 0.00, respectively.
Qwen3.5-Plus reaches the highest Hard-fault JRA of 0.29, followed by DeepSeek-V4-Flash at 0.22.
Strong results on Easy and Medium faults therefore do not imply comparable success on performance-related incidents.

Easy faults often provide direct clues, including image-pull events, scheduling failures, and explicit runtime errors.
These observations can narrow the diagnosis to a small set of components and fault types.
Medium faults require combining evidence across multiple components, Kubernetes resource objects, and system layers.
Although several models handle these investigations well, Hard faults require distinguishing mechanisms that can produce similar performance symptoms.
On OnlineBoutique, for example, resource saturation may result from a ChaosBlade-injected fault or from an application code defect.
The agent must therefore examine the relevant source code alongside telemetry to distinguish these causes.

TrainTicket's Hard cases involve ChaosBlade-induced performance faults, where Pods and services can appear healthy and the injection itself leaves no explicit diagnostic clues.
Locating these faults requires distinguishing the source of degradation from propagated symptoms across many services, a challenge central to telemetry-based RCA~\cite{OpenRCA2025ICLR,RCA2025Eval}.
Although \texttt{GetAlerts} provides threshold-based anomaly signals, these signals alone do not identify the root cause.
These characteristics help explain the low accuracy on TrainTicket Hard cases.
Despite strong results on simpler faults, the benchmark continues to distinguish model capabilities on Medium and especially Hard faults.

\begin{figure*}[t]
    \centering
    \includegraphics[width=\textwidth]{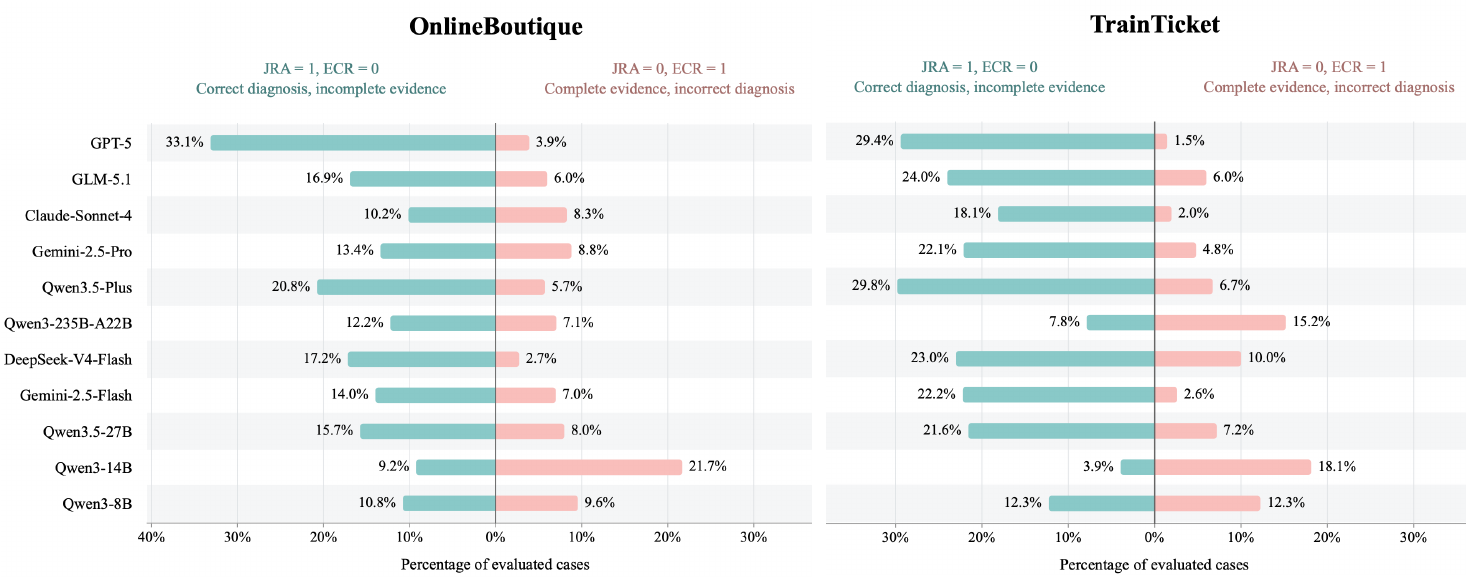}
    \vspace{-0.1in}
    \caption{Joint analysis of diagnostic correctness and evidence closure on OnlineBoutique and TrainTicket. Left bars show correct diagnoses with incomplete evidence (JRA=1, ECR=0), and right bars show complete evidence collection with incorrect diagnoses (JRA=0, ECR=1).}
    \label{fig:metrics}
    \vspace{-0.1in}
\end{figure*}

\subsection{RQ2: RCA Process Alignment}\label{sec:rq2}
We examine evidence completeness, acquisition order, and tool-use efficiency in the ReAct investigations reported in Table~\ref{tab:main}.

\textbf{(1) Correct diagnoses and complete evidence collection do not always coincide.}
Evidence closure varies substantially across models. On OnlineBoutique, Qwen3.5-27B reaches 0.72 JRA and 0.64 ECR, while Gemini-2.5-Pro reaches 0.67 JRA and 0.62 ECR. These agents complete the evidence requirements in a majority of their investigations. GPT-5 shows a larger difference, with 0.68 JRA and 0.39 ECR on OnlineBoutique; on TrainTicket, it reaches 0.68 JRA and 0.40 ECR. The aggregate scores establish that accuracy and closure can differ, but the joint results show what happens within individual investigations.

Figure~\ref{fig:metrics} separates the two directions of disagreement. For GPT-5, correct diagnoses with incomplete evidence account for 33.1\% of OnlineBoutique cases and 29.4\% of TrainTicket cases. The reverse pattern, complete evidence with an incorrect diagnosis, occurs in only 3.9\% and 1.5\%. Qwen3-14B shows the opposite balance: complete evidence with an incorrect diagnosis occurs in 21.7\% and 18.1\% of cases, exceeding its correct diagnoses with incomplete evidence (9.2\% and 3.9\%). Even for Qwen3.5-27B, whose aggregate JRA and ECR are close on OnlineBoutique, the two groups account for 15.7\% and 8.0\% of cases. A small difference between aggregate scores can therefore contain both kinds of disagreement.

Process scores clarify what an incomplete investigation has collected. GPT-5 reaches 0.70 MC but only 0.39 ECR on OnlineBoutique; Qwen3-14B reaches 0.66 MC and 0.39 ECR on TrainTicket. MC credits evidence acquired along an unfinished diagnostic path, whereas ECR requires a complete path. EOC adds the acquisition-order requirement: GPT-5's OnlineBoutique score falls from 0.70 MC to 0.62 EOC, and Qwen3-14B's TrainTicket score falls from 0.66 to 0.61. Some agents therefore stop before obtaining a required fact; others collect relevant facts before establishing their diagnostic prerequisites. Complete-but-incorrect investigations present a different problem: the required evidence was collected, but the final RCA was still wrong.

\textbf{(2) Auditing outcome-correct but unclosed trajectories.}
To verify whether unclosed trajectories reflect genuine investigative gaps rather than unannotated alternative paths in $\mathcal{D}_j^*$, we manually audited a random sample of 100 trajectories from GPT-5, DeepSeek-V4-Flash and Qwen3.5-27B. Overall, \textbf{95\%} of the audited cases stem from incomplete agent investigations. Specifically, Symptom-log termination without source tracing (73\%) accounts for the majority, where agents stopped immediately after observing a discriminative error log or event without tracing backward to verify the underlying configuration or resource state. While such log cues often correlate strongly with the fault label to yield a correct prediction, stopping at symptom logs leaves the concrete faulty artifact unverified for downstream remediation. Furthermore, Inconclusive evidence and heuristic guessing (22\%) involves cases where agents terminated with partial observations that remained logically ambiguous and could not rule out competing fault mechanisms. Only the remaining unannotated alternative paths (5\%) established the root cause through valid indirect observations outside our direct evidence bundles. Even if this 5\% margin were credited as fully closed, aggregate ECR scores would shift by less than 0.02, confirming that the outcome--process gap primarily reflects premature diagnostic termination rather than annotation false negatives.

\textbf{(3) Longer investigations do not necessarily produce more complete evidence.}
On OnlineBoutique, Qwen3-8B averages 15.78 steps with 0.34 EE and 0.19 ECR.
Its RAR of 0.45 shows that many calls repeat earlier tool-and-argument pairs without adding evidence or advancing a new milestone.
Qwen3-14B also conducts relatively long investigations, averaging 14.18 steps with 0.32 EE and 0.26 RAR, although it reaches a higher ECR of 0.55.
Repeated calls also occur on TrainTicket, where these models have RAR values of 0.22 and 0.31, respectively.
Continuing an investigation does not ensure that its remaining evidence gaps are addressed.

Shorter investigations can leave evidence incomplete for a different reason.
GPT-5 averages 5.56 steps on OnlineBoutique, with 0.64 EE and a near-zero RAR, but reaches only 0.39 ECR.
Many of its calls contribute evidence, yet the investigation often ends without completing the required checks.
GLM-5.1 averages 7.75 steps with 0.54 EE and 0.65 ECR, also with a near-zero RAR.
A higher fraction of evidence-contributing calls therefore does not necessarily mean that more investigations collect complete evidence.

These results show two different patterns of incomplete investigation.
Some agents finish after collecting useful but insufficient evidence.
Others continue calling tools but repeat checks that do not supply the missing facts.
Investigation length alone does not distinguish these behaviors or establish whether the diagnosis is fully supported.

\begin{table*}[t]
\centering
\caption{Comparison of experience reuse methods for RCA agents. The best results in each metric are bold.}
\vspace{-0.1in}
\label{tab:knowledge}
\scriptsize
\setlength{\tabcolsep}{2pt}
\resizebox{0.95\textwidth}{!}{
\begin{tabular}{llcccccccc|cccccccc}
\toprule
& & \multicolumn{8}{c}{\cellcolor[HTML]{DAE8FC}\textbf{OnlineBoutique}} & \multicolumn{8}{c}{\cellcolor[HTML]{FFFDD9}\textbf{TrainTicket}} \\
\cmidrule(lr){3-10}\cmidrule(lr){11-18}
\textbf{Model} & \textbf{Method} &
CA$\uparrow$ & FA$\uparrow$ & JRA$\uparrow$ & MC$\uparrow$ & EOC$\uparrow$ & ECR$\uparrow$ & EE$\uparrow$ & Steps$\downarrow$ &
CA$\uparrow$ & FA$\uparrow$ & JRA$\uparrow$ & MC$\uparrow$ & EOC$\uparrow$ & ECR$\uparrow$ & EE$\uparrow$ & Steps$\downarrow$ \\
\midrule
\multirow{3}{*}{DeepSeek-V4-Flash} & Base & 0.88 & 0.76 & 0.76 & 0.83 & 0.75 & 0.64 & \textbf{0.63} & 6.41 & \textbf{0.69} & \textbf{0.71} & \textbf{0.63} & \textbf{0.76} & 0.72 & \textbf{0.50} & \textbf{0.55} & 7.68 \\
 & ICL & 0.90 & \textbf{0.83} & \textbf{0.83} & \textbf{0.91} & \textbf{0.87} & \textbf{0.78} & \textbf{0.63} & 6.75 & 0.63 & 0.61 & 0.55 & \textbf{0.76} & \textbf{0.74} & 0.48 & 0.50 & 7.08 \\
 & SOP & \textbf{0.91} & \textbf{0.83} & 0.82 & 0.86 & 0.77 & 0.70 & 0.46 & \textbf{6.25} & 0.66 & 0.64 & 0.61 & 0.75 & 0.70 & 0.48 & 0.43 & \textbf{5.93} \\
\midrule
\multirow{3}{*}{Gemini-2.5-Flash} & Base & 0.82 & 0.67 & 0.65 & 0.79 & 0.75 & 0.60 & \textbf{0.58} & \textbf{6.49} & 0.60 & 0.55 & 0.53 & 0.57 & 0.54 & 0.30 & \textbf{0.53} & 4.91 \\
 & ICL & \textbf{0.84} & \textbf{0.74} & \textbf{0.74} & \textbf{0.84} & \textbf{0.81} & \textbf{0.69} & 0.54 & 6.88 & 0.59 & 0.60 & 0.58 & \textbf{0.73} & \textbf{0.71} & \textbf{0.48} & 0.49 & 6.87 \\
 & SOP & \textbf{0.84} & 0.71 & 0.67 & 0.81 & 0.76 & 0.64 & 0.41 & \textbf{6.49} & \textbf{0.68} & \textbf{0.64} & \textbf{0.61} & 0.69 & 0.64 & 0.39 & 0.45 & \textbf{4.86} \\
\midrule
\multirow{3}{*}{Qwen3.5-27B} & Base & 0.85 & 0.71 & 0.70 & 0.81 & 0.77 & 0.61 & 0.53 & \textbf{6.87} & 0.62 & \textbf{0.64} & 0.58 & \textbf{0.72} & \textbf{0.70} & 0.41 & \textbf{0.55} & 7.05 \\
 & ICL & \textbf{0.88} & \textbf{0.79} & \textbf{0.79} & \textbf{0.91} & \textbf{0.87} & \textbf{0.79} & \textbf{0.61} & 7.35 & 0.61 & 0.59 & 0.54 & 0.70 & 0.69 & \textbf{0.42} & 0.45 & 6.96 \\
 & SOP & 0.82 & 0.64 & 0.63 & 0.79 & 0.73 & 0.57 & 0.38 & 7.17 & \textbf{0.73} & 0.62 & \textbf{0.59} & 0.67 & 0.61 & 0.38 & 0.30 & \textbf{6.41} \\
\bottomrule
\end{tabular}
}
\end{table*}


\subsection{RQ3: Experience-Augmented RCA Agents}\label{sec:rq3}
In real cloud operations, engineers draw on prior incident traces, on-call notes, and runbooks to guide diagnosis.
We therefore examine whether reusable operational experience can improve RCA agents.
For each root cause, we partition the cases into disjoint training and test sets at an approximately 1:1 ratio.
The experience pool contains only successful GPT-5 trajectories from the training set, and Base, ICL, and SOP are evaluated on the same held-out test cases.

We compare two forms of experience reuse.
\emph{Trajectory-as-ICL}\cite{icl} retrieves three successful trajectories matched by the reported symptom, whereas \emph{Standard Operating Procedure (SOP)}\cite{FlowofAction2025WWW,gao2026rcaflow} distills recurring diagnostic actions and decision branches into category-level runbook guidance.
The target root cause is unknown during diagnosis, so neither method uses the ground-truth root-cause label to select experience for a test incident.
We select Gemini-2.5-Flash, Qwen3.5-27B, and DeepSeek-V4-Flash because they combine competitive base performance with relatively low inference cost and latency.
Table~\ref{tab:knowledge} reports the results.

\textbf{Historical experience can improve diagnosis, but its benefits vary across models, systems, and reuse methods.}
On OnlineBoutique, ICL improves both JRA and ECR for all three models.
Averaged across models, JRA increases from 0.70 to 0.79, and ECR increases from 0.62 to 0.75.
SOP produces smaller average gains, reaching 0.71 JRA and 0.64 ECR, with different effects across models.
It improves both metrics for DeepSeek-V4-Flash and Gemini-2.5-Flash, but reduces Qwen3.5-27B's JRA from 0.70 to 0.63 and ECR from 0.61 to 0.57.
Supplying successful historical experience therefore does not guarantee an improvement for every model.

The TrainTicket results are more mixed.
Average JRA changes from 0.58 for Base to 0.56 with ICL and 0.60 with SOP, while average ECR changes from 0.40 to 0.46 and 0.42, respectively.
ICL improves Gemini-2.5-Flash's JRA from 0.53 to 0.58 and ECR from 0.30 to 0.48, but reduces both metrics for DeepSeek-V4-Flash.
Accuracy and evidence completeness also respond differently to the same guidance.
For Gemini-2.5-Flash, SOP achieves higher JRA than ICL (0.61 versus 0.58), but lower ECR (0.39 versus 0.48).
The method that produces more correct diagnoses therefore need not produce more investigations with complete supporting evidence.

A practical difficulty is selecting relevant experience before the root cause is known.
Similar symptoms can arise from different mechanisms, so a retrieved successful trajectory may describe a different cause and omit checks needed for the current incident.
ICL provides detailed examples but is limited to three retrieved trajectories.
SOP covers recurring branches within a fault category, but the agent must still determine which branches apply.
These differences may contribute to the uneven gains across models and systems.

These results suggest that the effectiveness of experience reuse depends not only on the availability of prior successful diagnoses, but also on how that experience is represented, retrieved, and matched to the current incident. The incident reports, root-cause annotations, and snapshot-preserved system observations in \benchx provide a basis for studying more adaptive mechanisms for organizing and retrieving diagnostic experience.

\begin{table*}[h]
\centering
\caption{Comparison of agent architectures and harnesses using GLM-5.1. The best results in each metric are bold.}
\vspace{-0.1in}
\label{tab:harness}
\normalsize
\setlength{\tabcolsep}{2.6pt}
\resizebox{\textwidth}{!}{
\begin{tabular}{lccccccccc|ccccccccc}
\toprule
& \multicolumn{9}{c|}{\cellcolor[HTML]{DAE8FC}\textbf{OnlineBoutique}} & \multicolumn{9}{c}{\cellcolor[HTML]{FFFDD9}\textbf{TrainTicket}} \\
\cmidrule(lr){2-10}\cmidrule(lr){11-19}
\textbf{Method} & CA$\uparrow$ & FA$\uparrow$ & JRA$\uparrow$ & MC$\uparrow$ & EOC$\uparrow$ & ECR$\uparrow$ & EE$\uparrow$ & Steps$\downarrow$ & Tokens~(K)$\downarrow$ & CA$\uparrow$ & FA$\uparrow$ & JRA$\uparrow$ & MC$\uparrow$ & EOC$\uparrow$ & ECR$\uparrow$ & EE$\uparrow$ & Steps$\downarrow$ & Tokens~(K)$\downarrow$ \\
\midrule
\multicolumn{19}{l}{\textit{Agent architectures}} \\
ReAct & 0.93 & 0.84 & 0.81 & 0.86 & 0.76 & 0.68 & 0.57 & 7.44 & \textbf{55.13} & 0.62 & 0.60 & 0.55 & 0.65 & 0.63 & 0.38 & 0.35 & 10.17 & 129.46 \\
Plan-and-Execute & 0.90 & 0.77 & 0.76 & 0.84 & 0.75 & 0.64 & 0.52 & 7.98 & 60.47 & 0.72 & 0.65 & 0.62 & 0.77 & 0.73 & 0.50 & \textbf{0.48} & \textbf{9.85} & \textbf{126.77} \\
Reflection & 0.94 & 0.85 & 0.84 & 0.87 & 0.75 & 0.68 & \textbf{0.62} & \textbf{7.41} & 76.82 & 0.69 & 0.60 & 0.57 & 0.67 & 0.64 & 0.36 & 0.44 & 11.69 & 230.78 \\
\midrule
\multicolumn{19}{l}{\textit{Agent harnesses}} \\
Codex & \textbf{0.95} & \textbf{0.87} & \textbf{0.86} & \textbf{0.94} & 0.76 & \textbf{0.83} & 0.39 & 11.87 & 126.29 & \textbf{0.80} & \textbf{0.69} & \textbf{0.67} & \textbf{0.91} & \textbf{0.90} & \textbf{0.81} & 0.38 & 16.91 & 205.17 \\
OpenCode & 0.94 & 0.84 & 0.84 & 0.93 & \textbf{0.79} & 0.81 & 0.42 & 11.71 & 105.85 & 0.68 & 0.66 & 0.63 & 0.86 & 0.82 & 0.71 & 0.42 & 21.95 & 361.27 \\
\bottomrule
\end{tabular}
}
\par\smallskip
\begin{minipage}{\textwidth}
\scriptsize
\textit{Note:} Planning and control calls are excluded from step counts.
\end{minipage}
\end{table*}

\subsection{RQ4: Agent Architectures and Harnesses}
\label{sec:rq4}

The preceding experiments compare models and experience reuse within a ReAct-style loop.
Here, we hold the base model fixed at GLM-5.1 and compare five ways of organizing the investigation: ReAct\cite{yao2022react}, Plan-and-Execute\cite{wang2023plan}, Reflection\cite{shinn2023reflexion}, Codex, and OpenCode.
Plan-and-Execute uses a planner to select diagnostic subgoals and a ReAct executor to investigate them, with replanning based on the accumulated evidence until the planner submits the final RCA.
Reflection reviews a candidate diagnosis against the collected evidence and permits further investigation and answer refinement.
We also evaluate two coding-agent harnesses, Codex CLI and OpenCode,\footnote{We use Codex CLI v0.139.0 and OpenCode v1.18.31.}
which provide tool-mediated execution, context management, and iterative investigation workflows.
We connect these harnesses to the benchmark's diagnostic tools through Model Context Protocol  (MCP).

In this RQ, we increase the diagnostic tool-call budget from 20 in the main experiments to 50 to accommodate the additional diagnostic checks introduced by planning, orchestration, and recovery workflows~\cite{hao2026cocoabench}, and apply the same budget to all five methods.
Planning and control calls are excluded from this budget.
Table~\ref{tab:harness} reports results over all cases.
To compare the computational costs of these agent implementations, we additionally report mean end-to-end input and output token consumption per case, expressed in thousands (K).

\textbf{(1) Similar diagnostic accuracy can accompany different levels of evidence completeness.}
On OnlineBoutique, Reflection and OpenCode both achieve 0.84 JRA, but their ECR values are 0.68 and 0.81, respectively.
On TrainTicket, Plan-and-Execute and OpenCode achieve similar JRA values of 0.62 and 0.63, while their ECR values differ more substantially, at 0.50 and 0.71.
Thus, agents that correctly diagnose similar proportions of cases can differ in how often they collect complete supporting evidence.

Codex achieves the highest JRA and ECR on both systems in this experiment, reaching 0.86 and 0.83 on OnlineBoutique, and 0.67 and 0.81 on TrainTicket.
Its investigations also use more diagnostic steps than ReAct, Plan-and-Execute, and Reflection.
The results indicate that the organization and continuation of an investigation affect how much evidence an agent collects, even when the underlying model and diagnostic tool-call budget are held fixed.

\textbf{(2) More complete investigations consume more tokens, but higher expenditure does not guarantee better results.}
On OnlineBoutique, Codex uses 126.29K tokens per case, approximately 2.29 times ReAct's 55.13K.
Its ECR increases from ReAct's 0.68 to 0.83, while JRA increases from 0.81 to 0.86.
The additional computation accompanies improvements in both metrics, with a larger increase in evidence closure.

On TrainTicket, however, OpenCode uses 21.95 steps and 361.27K tokens per case, compared with Codex's 16.91 steps and 205.17K tokens, yet achieves lower JRA (0.63 versus 0.67) and ECR (0.71 versus 0.81).
Codex therefore achieves both higher diagnostic accuracy and more complete evidence collection at a lower token cost in this comparison.

\subsection{Analysis of RCA Agent Diagnostic Processes}
We inspect recorded agent trajectories to examine how evidence acquisition and interpretation shape diagnostic outcomes.
The analysis covers both incorrect diagnoses and correct diagnoses with incomplete evidence collection.
The selected trajectories illustrate representative behaviors rather than estimate their prevalence.
\subsubsection{Failure Modes in Incorrect Diagnoses}
To understand agent errors beyond aggregate scores, we first collect failure cases with $\mathrm{JRA}=0$ from the 11 evaluated LLMs. From this failure pool, we qualitatively inspect a purposive sample of trajectories covering both benchmark systems, all eight fault categories, and models with different capability levels. This exploratory analysis characterizes representative and potentially overlapping behaviors rather than estimating their prevalence or defining an exhaustive taxonomy. We organize the observed failures around four broad forms of diagnostic breakdown.

\textbf{(1) Evidence-grounding failures.}
Agents sometimes retrieve useful evidence but interpret it incorrectly or assign it insufficient weight. For example, startup events may already report \texttt{manifest unknown}, \texttt{403 Forbidden}, or a missing secret, yet the agent continues checking peripheral resources. Agents can also anchor on salient downstream symptoms, such as frontend 503 responses, high latency, or a prominent CPU alert, and mistake the noisiest component for the causal source. Conversely, missing objects and telemetry can be overlooked: an empty endpoint set, an absent pod, or the lack of logs from a failed component is sometimes treated as no evidence rather than as a diagnostic signal. Such misinterpretations can place the investigation in the wrong fault family, for example by explaining a code defect as OOM, network delay, or service misconfiguration.

\textbf{(2) Search-control failures.}
Agents do not always adapt their investigation to the evidence already established. Some commit to a plausible hypothesis after inspecting one abnormal pod or service and stop before checking competing explanations. Others fail to redirect the search when an initial path is weakened; for instance, after confirming normal service connectivity, they may not move to gateway routing or configuration checks. The opposite behavior also occurs: even after observing decisive evidence such as \texttt{replicas=0}, an inactive \texttt{kube-proxy}, or an explicit database access error, an agent may continue issuing low-value or repeated queries. These trajectories reflect weak control over hypothesis revision, diagnostic branching, and stopping.

\textbf{(3) Target-resolution failures.}
Relevant evidence does not necessarily lead agents to the correct diagnostic target. In TrainTicket, database errors observed by application services may be attributed to those services instead of the shared \texttt{tsdb-mysql} dependency; similarly, gateway routing faults may be assigned to affected business services. Agents also confuse closely related mechanisms, such as image-reference and registry-DNS failures, probe port and protocol errors, or adjacent code-defect subtypes. Codedefect cases further require connecting service-level alerts, HTTP errors, and abnormal metrics to source-level behavior. Agents may inspect relevant telemetry or code artifacts but still explain a busy loop, artificial delay, or excessive file operation as resource saturation or a performance fault.

\textbf{(4) Tool-grounding failures.}
Failures also arise at the interface between reasoning and execution. Invalid parameters, an incorrect object or scope, or an action inconsistent with the intended check can prevent the required evidence from being obtained. More subtly, a syntactically valid but mis-specified query can create misleading evidence. For example, probing service connectivity with the wrong port produces a genuine connection failure that the agent may then treat as a production symptom and use to support an incorrect diagnosis.

Our process-oriented analysis reveals that agent failures often develop before the final prediction, through breakdowns in evidence-grounded investigation and reasoning. These observations show that improving current RCA agents requires not only better final predictions, but also more reliable and interpretable diagnostic trajectories.
\subsubsection{Case Studies: Evidence Gaps in Correct Diagnoses}
\label{sec:casestudy}
We examine three representative investigations in which the agent correctly identifies both the faulty component and root cause but receives $\mathrm{ECR}=0$. For each case, we compare the collected observations with the evidence available in the frozen snapshot to identify what remains unverified. These cases illustrate recurring forms of evidence incompleteness rather than model-specific failures or their prevalence.

\textbf{(1) Missing backend-label evidence.}
The incomplete comparison illustrated in Fig.~\ref{fig:case-score} also appears in actual agent trajectories. In OnlineBoutique \texttt{service/38}, a representative run observes that \texttt{shippingservice} has no endpoints and selects Pods with \texttt{app=shipping\_service}. The agent correctly reports a selector mismatch but never retrieves the backend Pod labels or equivalent configuration evidence. The frozen snapshot shows that the backend Pod is actually labeled \texttt{app=shippingservice}, but this fact is absent from the collected observations. The diagnosis is therefore correct, yet the selector--label mismatch itself is never directly verified, resulting in $\mathrm{MC}=0.75$ and $\mathrm{ECR}=0$.

\textbf{(2) Testing a connection without checking the exposed port.}
In TrainTicket \texttt{runtime/58}, a representative trajectory correctly diagnoses an invalid MySQL port in \texttt{ts-auth-service}. The application configuration points to \texttt{tsdb-mysql-leader:34001}; the agent observes connection-refused errors and confirms that this address is unreachable. However, it never inspects the database Service or Endpoints to determine the port actually exposed. The frozen snapshot shows that the Service exposes port \texttt{3306}, but this fact is absent from the collected observations. The connectivity test therefore confirms failure at the configured address without establishing the port mismatch itself, leaving the investigation at $\mathrm{MC}=0.75$ and $\mathrm{ECR}=0$.

\textbf{(3) Attributing latency to code without inspecting the implementation.}
In OnlineBoutique \texttt{codedefect/69}, The agent correctly reports an artificial delay in \texttt{productcatalogservice}. It retrieves alerts, Pod status, error logs, and application YAML. The alerts show that the service's p90 latency increases from 2.13\,ms to 4003.98\,ms, while the error-log query returns no relevant errors. However, the agent never invokes \texttt{ListCodeFiles} or \texttt{GetSourceCode}. The frozen snapshot contains an explicit three-second sleep in \texttt{ListProducts}, but this implementation evidence is never retrieved. The trajectory therefore establishes the latency symptom without verifying the code-level mechanism, resulting in $\mathrm{MC}=1/3$ and $\mathrm{ECR}=0$.

Across the three cases, the submitted RCA labels are correct, yet key supporting facts remain uncollected. These examples demonstrate that outcome correctness alone cannot distinguish a fully grounded diagnosis from one inferred from partial but salient evidence.

\vspace{-0.1cm}
\section{Discussion}

\subsection{Implications and Research Utility} \label{sec:significance}

\benchx is an evaluation infrastructure for interactive and evidence-grounded cloud RCA.
Its contribution extends beyond a collection of cloud fault cases by making both the final diagnosis and its supporting investigation controlled and measurable.

\textbf{Controlled experiments on cloud diagnosis.}
By replaying the same runtime-verified incident, \benchx allows researchers to compare RCA agents or vary one agent component while holding the available fault evidence fixed.
For example, planners, memory mechanisms, topology and knowledge retrieval, and tool-selection policies can be compared using the same user query and incident state.
Differences in diagnostic correctness, milestone coverage, evidence-order consistency, and investigation efficiency can therefore be attributed more directly to agent design rather than to a different realization of the fault.
Unlike static telemetry inputs, the replay interfaces preserve the agent's decisions about which diagnostic tool, component, and evidence source to inspect.

\textbf{Process-level analysis of RCA agents.}
The diagnostic evidence graph aligns different tool-mediated investigations according to the causal facts they establish.
For a failure that propagates across services or platform resources, it can show whether an agent observes the reported manifestation but fails to traverse the relevant dependency, reaches the faulty component without confirming its mechanism, or concludes before acquiring sufficient evidence.
Alternative evidence bundles and partial-order dependencies allow these failures to be analyzed without requiring agents to reproduce a canonical command sequence.
This enables targeted studies of whether topology information improves propagation tracing, retrieved operational knowledge improves mechanism confirmation, or memory and planning reduce repeated or causally premature investigation.

\textbf{Reusable diagnostic experience.}
Trajectories produced on \benchx can be aligned to the same milestones even when agents follow different investigation paths.
This alignment makes successful and failed strategies comparable across models, fault types, and application and platform layers, providing a basis for studying how diagnostic experience transfers across incidents.
Although this work focuses on evaluation, the resulting trajectories can also support future research on experience retrieval, diagnostic skill distillation, and process-aware training for cloud RCA agents.

\vspace{-0.1cm}
\subsection{Future Research Opportunities} \label{sec:future_work}

\textbf{State-adaptive evidence acquisition.}
Current RCA agents often rely on generic investigation routines that do not adapt sufficiently to the evidence already established.
Future work can study policies that select the next tool and target according to the grounded milestones, unresolved diagnostic alternatives, and expected information gain.
An important question is whether topology-aware planning and uncertainty-aware stopping can improve evidential completeness while avoiding premature conclusions and unnecessary inspection.

\textbf{Transferable diagnostic skills.}
Cloud incidents differ in their components and topologies, but often require recurring diagnostic operations such as dependency traversal, failure-boundary localization, and mechanism confirmation.
Aligning trajectories through diagnostic milestones provides a basis for learning or retrieving skills that are invariant to component identifiers and exact command syntax.
Future studies should examine whether experience acquired from one workload or fault type improves investigation of unseen incidents without relying on root-cause labels or memorized case-specific paths.

\textbf{RCA under incomplete and interacting evidence.}
The current benchmark focuses on individually verified faults whose required evidence is captured in the verified incident state or its associated case-specific incident interval.
Operational incidents may instead involve concurrent faults, delayed effects, changing system states, and missing or contradictory observations.
Extending the construction pipeline to such settings would enable research on maintaining competing hypotheses, distinguishing root causes from propagated symptoms, and expressing uncertainty when the available evidence does not support a unique diagnosis.
Such extensions should preserve matched evaluation conditions and auditable evidence requirements rather than introducing complexity at the expense of comparability.

\subsection{Threats to Validity}
\textbf{External Validity.}
\benchx focuses on Kubernetes-based cloud-native systems, which may limit generalization to other architectures and operational environments.
We mitigate this threat by covering \faulttype fault types drawn from multiple sources, two workloads with different service scales, both application and platform layers, and retaining only runtime-verified cases.
Nevertheless, broader validation requires additional workloads, deployment configurations, and real-world incidents.
Each benchmark release also reflects the Kubernetes version used during fault realization and snapshot capture.
Most supported diagnostic interfaces rely on stable API resources and long-standing troubleshooting operations governed by Kubernetes compatibility and deprecation policies~\cite{k8s_deprecation_policy}, which reduces drift under routine upgrades.
When a Kubernetes update changes a fault manifestation or diagnostic interface, the construction pipeline can re-execute the affected cases and recapture their snapshots as a revalidated, versioned release.

\textbf{Internal Validity.}
Model comparisons may be affected by the agent scaffold, prompting choices, inference nondeterminism, and differences in prior exposure to Kubernetes knowledge.
We mitigate these factors by using the same ReAct framework, prompt structure, tool schemas, step budget, temperature setting, and snapshot evidence across models.
Because process credit requires case-specific observations acquired through interaction, prior domain knowledge alone cannot obtain a complete process score.

\textbf{Construct Validity.}
\benchx evaluates diagnosis from evidence preserved in the verified incident state and its associated case-specific incident interval rather than from arbitrary system history or subsequent live state transitions.
We mitigate this limitation by covering the principal evidence channels used in cloud-native diagnosis and retaining only cases whose required milestones are observable and replay-validated.
Faults whose diagnosis requires long-term historical baselines, evidence preceding baseline confirmation, or observations produced only after verification are outside the current benchmark scope.

Furthermore, our process metrics require positive observational verification of the underlying faulty state and mechanism. In principle, an agent may predict the correct label by relying on salient downstream error logs or indirect elimination without directly inspecting the faulty configuration or resource. Under our scoring, such trajectories receive partial Milestone Coverage (MC) credit for the collected clues but $\text{ECR}=0$. We intentionally maintain this requirement because actionable production RCA requires directly verifying the root-cause artifact before remediation, rather than relying on unverified symptom correlations or elimination alone.

\vspace{-0.1cm}
\section{Related Work}



\textbf{Evolution of RCA.}
Traditional AIOps RCA formulates diagnosis as discriminative localization over metrics, logs, and traces, ranking faulty components with causal or graph models~\cite{Dejavu2022FSE,yu2021microrank,nezha2023fse,changerca2024fse,MicroSketch2022,Swisslog2023TPDS,Microservie2018TSE,FaaSRCA2024ISSRE,lee2023eadro}.
LLM-based systems then treat RCA as generative analysis over a provided incident context, often retrieving operational documents to produce explanations~\cite{X-Lifecycle2024FSE,Incident2023ICSE,RCACopilot,COCA,SynergyRCA}.
Recent agentic workflows go further: they select diagnostic tools, interpret returned observations, and revise hypotheses through multi-step interaction with cloud and related operational environments~\cite{agent2024IJCAI,DBAIOps2025VLDB,StepFly2025,RCAgent2024CIKM,FlowofAction2025WWW,AidAI25FSE,GALA2025,LLMGuard}.
This shift makes interactive evidence acquisition central to RCA and motivates a new class of benchmarks that can evaluate not only final root-cause labels, but also tool-mediated investigation under controlled, comparable incident conditions.

\textbf{Benchmarking in AIOps.}
Existing AIOps benchmarks only partially meet this need.
Static telemetry benchmarks provide deterministic offline artifacts for detection or localization, but agents cannot investigate through operational interfaces~\cite{LogEval2025,Loghub2023ISSRE,RCA2025Eval,nezha2023fse,TimeSeriesBench2024ISSRE,MTSAD2024TimeSeries,aiops2020,aiops2021,lee2023eadro}.
OpenRCA allows code-based queries over supplied files, which improves inspectability without exposing system-facing SRE tools~\cite{OpenRCA2025ICLR}.
Live-environment benchmarks restore standard tool use on real services, but fault manifestations can vary across runs, which complicates controlled comparison and fine-grained process scoring~\cite{chen2025aiopslab,ITBench2025ICML,AIOpsArena25SANER,clark2026sregym}.
Across both paradigms, evaluation remains largely outcome-oriented.
As summarized in Table~\ref{tab:comparison}, \benchx targets the missing combination: snapshot-backed replay preserves system-facing investigation under matched incident observations, and diagnostic evidence graphs score whether required facts are established in an admissible order from tool observations, rather than only whether the final root-cause label is correct or whether the agent reproduces a single expert command sequence.

\textbf{Trajectory and Process Evaluation of LLM Agents.} 
Recent studies have explored evaluating intermediate agent trajectories across reasoning, web, and desktop environments. In mathematical reasoning, process supervision~\cite{prm} trains reward models to score the validity of individual derivation steps. For interactive task execution, AgentBoard~\cite{AgentBoard} tracks fine-grained progress through environment states and subgoal completion, while WebCanvas~\cite{WebCanvas} evaluates web agents through critical intermediate actions and states. CocoaBench~\cite{hao2026cocoabench} evaluates long-horizon digital-agent workflows using task-specific automatic evaluators, while OSWorld-Pro~\cite{wang2026osworld} decomposes computer-use tasks into sequentially dependent subgoals and employs LLM-based judges~\cite{llmjudge} for process-level assessment. However, these paradigms primarily evaluate reasoning-step correctness, task progress, or observable execution outcomes, rather than diagnostic evidence acquisition. In contrast, cloud-native RCA is a read-only diagnostic probing task: agents query a distributed system without mutating its state, and the same diagnostic fact can be uncovered through multiple tool channels, query scopes, and non-linear investigation orders. While OpenRCA 2.0~\cite{openrca2} introduced PAVE to reconstruct fault-propagation paths using fault-injection information and telemetry, Cloud-OpsBench formalizes deterministic process evaluation over interactive tool observations, coupling conjunctive and disjunctive completion logic and prerequisite dependencies with exhaustively enumerated snapshot evidence bundles.

\vspace{-0.1in}
\section{Conclusion}
This paper presents \benchx, an evaluation infrastructure for interactive and evidence-grounded cloud RCA.
It combines runtime-verified fault realization with snapshot-backed replay, allowing agents to investigate matched incident observations through standard diagnostic interfaces while independently selecting their tools and targets.
Each case pairs an outcome label with a diagnostic evidence graph, enabling evaluation of root-cause correctness, milestone coverage, evidence-order consistency, and investigation efficiency without prescribing a canonical command sequence.
With \num cases spanning \faulttype fault types across application and Kubernetes platform layers, \benchx provides a controlled basis for comparing RCA agents and analyzing where their investigations fail in evidence acquisition, dependency traversal, mechanism confirmation, and final diagnosis.


\bibliographystyle{ACM-Reference-Format}
\bibliography{ref}

\end{document}